\documentclass[]{tCPH2e}

\begin{document}

\def\a{\alpha}
\def\b{\beta}
\def\c{\chi}
\def\d{\delta}
\def\e{\epsilon}
\def\f{\phi}
\def\g{\gamma}
\def\h{\eta}
\def\i{\iota}
\def\j{\psi}
\def\k{\kappa}
\def\l{\lambda}
\def\m{\mu}
\def\n{\nu}
\def\o{\omega}
\def\p{\pi}
\def\q{\theta}
\def\r{\rho}
\def\s{\sigma}
\def\t{\tau}
\def\u{\upsilon}
\def\x{\xi}
\def\z{\zeta}
\def\D{\Delta}
\def\F{\Phi}
\def\G{\Gamma}
\def\J{\Psi}
\def\L{\Lambda}
\def\O{\Omega}
\def\P{\Pi}
\def\Q{\Theta}
\def\S{\Sigma}
\def\U{\Upsilon}
\def\X{\Xi}

\def\ve{\varepsilon}
\def\vf{\varphi}
\def\vr{\varrho}
\def\vs{\varsigma}
\def\vq{\vartheta}
\def\dg{\dagger}                                     
\def\ddg{\ddagger}                                   
\def\wt#1{\widetilde{#1}}                    
\def\mt{\widetilde{m}_1}
\def\mti{\widetilde{m}_i}
\def\rt{\widetilde{r}_1}
\def\mtt{\widetilde{m}_2}
\def\mttt{\widetilde{m}_3}
\def\rtt{\widetilde{r}_2}
\def\mb{\overline{m}}
\def\VEV#1{\left\langle #1\right\rangle}        
\def\be{\begin{equation}}
\def\ee{\end{equation}}
\def\ds{\displaystyle}
\def\ra{\rightarrow}

\def\bea{\begin{eqnarray}}
\def\eea{\end{eqnarray}}
\def\NO{\nonumber}
\def\Bar#1{\overline{#1}}

\doi{10.1080/0010751YYxxxxxxxx}
 \issn{1366-5812}
\issnp{0010-7514}

\jvol{00} \jnum{00} \jyear{2012}  \jmonth{June}
\setcounter{footnote}{0}

\markboth{}{}

\articletype{}

\title{An introduction to leptogenesis and neutrino properties}

\author{Pasquale Di Bari\\
\vspace{6pt}  
{\em{School of Physics and Astronomy, University of Southampton, Southampton SO17 1BJ, UK}};\\
\vspace{6pt}\received{June 2012} 
}

\maketitle

\begin{abstract}
This is an introductory review of the main features of leptogenesis, 
one of the most attractive models of baryogenesis 
for the explanation of the matter-antimatter asymmetry of the Universe. 
The calculation of the  asymmetry in leptogenesis is intimately related to neutrino
properties so that leptogenesis is also an important phenomenological tool
to test the see-saw mechanism for the generation of neutrino masses and mixing and
the underlying theory beyond the Standard Model.
\bigskip
\bigskip

\end{abstract}

\section{The double side of leptogenesis}

It was first pointed out by Andrej Sakharov in 1967 \cite{sakharov}
that the non observation of primordial antimatter in the observable Universe
could be related to fundamental properties of particle physics and in particular
to the violation of a particular symmetry, called $C\!P$, that was 
discovered  just three years before the Sakharov paper in the decays of 
newly discovered particles  called $K$ mesons \cite{cpv}.
The  $C\!P$ transformation is the combination of the   
transformation of parity $P$, corresponding to flipping the sign of all space coordinates,
and of charge conjugation $C$, corresponding to flip the sign of all charges of the elementary particles,
including the electric charge.  The fact that the laws of physics are not invariant under $C\!P$
transformation, implies that matter and antimatter can behave differently in elementary processes.

Today the prescient idea of Sakharov is supported by a host of cosmological observations
that show how the matter-antimatter asymmetry of the Universe has to be explained
in terms of a dynamical generation mechanism, what is called a model of baryogenesis, 
incorporating $C\!P$ violation. 
At the same time it has been realised that a successful model of baryogenesis cannot be
attained within the current Standard Model (SM) of particle physics and it has therefore
to be regarded as a necessity of extending the SM with a model beyond the SM, or, more shortly, 
with some `new physics'.  
 
The discovery of neutrino masses and mixing in neutrino oscillation
experiments in 1998 \cite{SuperK}, has for the first time shown directly,
in particle physics experiments, that the SM is indeed incomplete, since it strictly 
predicts that neutrinos are massless and, therefore, cannot oscillate. 
 
Therefore, this discovery has greatly increased the interest in 
a mechanism of baryogenesis called  leptogenesis \cite{fy},  a model of 
baryogenesis that is a cosmological consequence of the most 
popular way to extend the SM in order to explain
why neutrinos are massive but at the same time much lighter than all
other fermions: the see-saw mechanism \cite{see-saw}. 

In this way leptogenesis  realises a highly non trivial link
between two completely independent experimental observations:   
the absence of primordial antimatter in the observable Universe and the 
observation that neutrinos mix and (therefore)  have masses.
Leptogenesis has, therefore, a naturally built-in double sided nature.
On one side, it describes a very early stage in the history of the Universe 
characterised by temperatures $T_{\rm lep} \gtrsim 100\,{\rm GeV}$,
much higher than those probed by Big Bang Nucleosynthesis ($T_{BBN} \sim 1\,{\rm MeV}$)
\footnote{Throughout the review, we adopt the natural system with $\hbar = c = k_B = 1$
and therefore temperatures are expressed in electronvolts, the usual
unit of energy in particle and early Universe physics.}, 
and on another  side it complements low energy neutrino experiments providing a completely independent phenomenological tool 
to test models of new physics embedding the see-saw mechanism. 

Before reviewing the main features and results in leptogenesis,
we will first need to discuss, in the next section, the cosmological framework justifying the necessity
of a model of baryogenesis and then, in Section 3, the main experimental results in the study
of neutrino masses and mixing parameters
and the main features of the see-saw mechanism. Neutrino experiments
and see-saw mechanism together  suggest leptogenesis to be today probably the most 
plausible candidate 
for the explanation of the matter-anti matter asymmetry of the Universe.

\section{The cosmological framework}

\subsection{The $\L$CDM model: a missed opportunity for a cosmological role of neutrino masses?}
\setcounter{footnote}{1}

In the hot Big Bang model, the history of the Universe has two quite well distinguished stages. 
In a first hot phase, the early Universe,  matter and radiation were coupled
and the growth of baryonic matter perturbations was inhibited. In this stage
matter was in the form of a plasma and properties of elementary particles
were crucial in determining the evolution of the Universe.

After matter-radiation decoupling occurring during the so called
re-combination epoch
\footnote{Within our current cosmological model, electrons combine with nuclei to form
atoms for the first time during  `recombination', roughly 300,000 years after the Big Bang. 
Therefore, this should be more
correctly called  `first combination epoch'. To make things even more confusing notice
that  there was even a second period of ionisation followed by
a second combination epoch, approximately half billion of years after the Big Bang, 
that should, more legitimately, be called recombination. 
However, for historical reasons, the first combination cosmological epoch 
is universally known as `recombination', while the second epoch is 
called `re-ionisation'.},
when electrons combined with protons and 
Helium-4 nuclei to form atoms, baryonic  matter perturbations could grow 
quite quickly under the action of Dark Matter inhomogeneities, 
forming the large-scale structure that we observe. 
	
With the astonishing progress in observational cosmology during the last 15 years,
we have today quite a robust minimal cosmological model, the so called 
`$\Lambda$-Cold Dark Matter' ($\L$CDM) model
sometime popularly dubbed as the `vanilla model'.
The $\L$CDM model is very successful in explaining all current cosmological observations 
and its parameters are currently measured with a precision better than $\sim 10\%$. 

The $\L$CDM model relies on General Relativity, the Einstein's theory of gravity,
for a description of the gravitational interactions on cosmic scales.  It belongs
to the class of Friedmann cosmological models based on the assumption of the
homogeneity and isotropy of the Universe.  
In this case the space-time geometry is conveniently described in the so called comoving system
by the Friedmann-Robertson-Walker metric in terms of just  one time-dependent 
parameter: the scale factor.  The distances among objects 
at rest with respect to the comoving system are just all proportional to the
scale factor, as the distances of points on the surface of an inflating balloon.   
The expansion of the Universe is then described in terms of the scale factor 
time dependence that can be worked out as a solution of the Friedmann equations. 

However, a solution of the Friedmann equations also requires the knowledge of the energy-matter
content of the Universe. In this respect, the $\L$CDM model also belongs to a sub-class of Friedmann
models, the Lemaitre models, where the energy-matter content is described by
an admixture of three different forms of fluids: 
i) matter, the non-relativistic component where energy
is dominated by the mass term; 
ii) radiation, the ultra-relativistic component
where  energy is dominated by the kinetic energy term (in the case of photons this is exactly true);
iii) vacuum energy density and/or a cosmological constant term. 

The latter, if positive, can be responsible for a repulsive behaviour of gravity and it was first introduced
by Einstein in 1917 in the equations bearing his name in order to balance the attractive action of matter
and obtain a static Universe. 
At that time there was no evidence for an expansion of the Universe and a static model 
represented the most attractive solution. However,  with the discovery of the 
expansion of the Universe by Edwin Hubble in 1929, it became clear that a successful
cosmological model should contain a time variation of the scale factor. 
The cosmological constant became for a long time
a mere theoretical possibility invoked, from time to time, to solve different observational
problems without however any conclusive evidence. 
Lemaitre studied in great details these models containing an additional cosmological constant term. 

Today, thanks to  the precision and accuracy of the current cosmological observations, 
we have a very strong indication that indeed the expansion
of the Universe is described by a Lemaitre model with a positive cosmological constant
whose action is completely equivalent to have a non-vanishing vacuum energy density. 
The necessity of a positive cosmological constant comes from
the observed acceleration of the Universe at the present time, first discovered 
from the determination  of the Supernovae type Ia redshifts versus luminosity distances
 relation, a discovery that has been awarded the 2011 Nobel Prize for Physics \cite{SNIa}.
The presence of a cosmological constant, indicated with the symbol $\L$,
as in the Einstein original paper, is therefore an important  
distinguishing  feature of the $\L$CDM model, as clearly indicated by the name. 
So far any attempt
to reproduce the acceleration of the Universe with an alternative explanation either
failed or it is indistinguishable from $\L$ in current observations and it therefore
represents an unnecessary complication of the $\L$CDM model that is in this way
the minimal model able to reproduce all observations. 

\setcounter{footnote}{2}

In Friedmann models, the space-time geometry can be of three types, closed, open or flat
depending whether the present value of the energy density parameter, $\O_0 \equiv \rho_0/\rho_{\rm c,0}$,
is respectively bigger, smaller or equal to $1$, where
$\rho_0$ is the total energy density and  $\rho_{\rm c,0}$ is 
the so called critical energy density,  both calculated at the present time
\footnote{The critical energy density is, therefore, that particular value of the
total energy density corresponding to a flat geometry and is defined as 
$\rho_{c,0}\equiv 3\,H_0^2/(8\pi\,G)$, where $H_0$ is the Hubble constant
and $G$ is Newton's constant.}.  

The cosmological observations strongly indicate that the current geometry of the Universe
is very close to be flat, a point to which we shall soon return. 
 In the $\L$CDM model, the flatness of the Universe
is set up to be exactly zero,  another important distinguishing feature, that has a very
robust theoretical justification, as we will discuss.  

The current cosmological observations are also able to determine quite precisely
the values of the different contributions to the total energy density 
from the three different fluids in the $\L$CDM model: radiation
contributes at the present time only with a tiny $0.1\%$, matter gives a more significant $26.5 \%$ contribution,
while the dominant remaining $73.5\%$ contribution is in the 
form of a cosmological constant and/or vacuum energy density. This particular
combination of values,  a sort of cosmological recipe, 
reproduces very well all cosmological observations and in particular the 
above-mentioned acceleration of the Universe at the present time. 

A fundamental ingredient of the $\L$CDM model  is the existence
of a very early stage in the history of the Universe called inflation \cite{inflation}, characterised by
a super-luminal expansion, that was able to bring a microscopic sub-atomic portion of space
to have a macroscopic size corresponding today to our observable Universe 
where a homogeneous, isotropic and a flat space-time geometry 
holds with very good approximation. 
In this way homogeneity, isotropy and flatness of the Universe have not to be
postulated but are instead a natural result of the inflationary stage.
However, inflation also predicts, at the end of the Inflationary stage, 
the presence of primordial perturbations that acted as seeds 
for the formation of Galaxies, clusters of Galaxies and super-clusters of Galaxies:
the so called large-scale structure of the Universe.  
The same primordial perturbations are also responsible for the 
observed Cosmic Microwave Background (CMB) temperature anisotropies and for their properties. 
In particular for the explanation of the so called acoustic peaks in the angular power
spectrum of the CMB temperature anisotropies. The acoustic peaks originate from
the compression and rarefaction of the coupled baryon-photon fluid at the time
of recombination. The angular power spectrum at a value of the multipole
number ${\ell}$ is determined by the presence of 
anisotropies of angular size $\Delta\theta \simeq 180^{\circ}/{\ell}$. The existence and 
position of the first peak at a multipole number ${\ell}_1 \simeq 200$, 
corresponds to regions of the early Universe with a size given by the distance
that sound could have traveled within the recombination time, the 
so called {\em sound horizon}.  The angular size of such a region depends also 
on the geometry of the Universe and it can be proven that in general the position
of the first peak would be ${\ell}_1\simeq 200/\sqrt{\Omega_0}$. The experimental
observation that ${\ell}_1 \simeq 200$ provides therefore the strongest indication
that $\Omega_0 \simeq 1$, confirming, within current experimental errors,  
the expectation of a flat Universe geometry coming from Inflation. 
The acoustic peaks are probably today the most important source of experimental 
information on the cosmological parameters supporting the $\L$CDM model.  

From this precise and accurate determination of the cosmological parameters 
of the $\L$CDM model,  we also know that the early Universe
stage lasted about 450,000 years, only a short time compared to the 
total age of the Universe $t_0$ that in the $\L$CDM turns out to be about 
13.75 billions of years. 

\subsection{Neutrinos in the early Universe}

The existence of a first hot stage is not only fundamental to understand the existence 
and the properties of the Cosmic Microwave Background radiation (CMB) 
and the nuclear composition of the Universe with Big Bang Nucleosynthesis (BBN) 
but it also seems to enclose the  secrets for the solution of those 
cosmological puzzles of the  $\Lambda$CDM model that strongly hint at New Physics. 
These include: i) the existence of Dark Matter, that has a crucial role in making possible the
quick formation of galaxies after the matter-radiation decoupling; ii) the observed matter-antimatter
asymmetry of the Universe; iii) the necessity of an inflationary stage and iv) the presence
of the mysterious form of energy, currently indistinguishable from a cosmological constant, 
that is driving the acceleration of the Universe at the present time.

\setcounter{footnote}{3}

Interestingly recent CMB observations seem also to hint at the presence of some extra-radiation
during recombination \cite{WMAP7}. Indeed within the Standard Model of particle physics (SM)
and from the existing upper bounds on neutrino masses, that we will discuss in a moment, we know that
the only particle species that can contribute to the radiation component
during the recombination time, to which the observed CMB properties can be ascribed, 
are photons and three known species of neutrinos in thermal equilibrium 
\footnote{In this sense
photons and neutrinos are said to be a thermal radiation component
within the $\L$CDM model.  There are also models where
a non-thermal radiation component can be present.}. Current observations seem to suggest the presence 
of an additional radiation component equivalent approximately to one additional  neutrino 
species in thermal equilibrium. The results from the PLANCK satellite, launched in June 2010 and currently 
taking data able to yield the most accurate CMB anisotropies maps, are expected to be published in
January 2013 and will be able to shed a light on this current anomaly, 
since it will be able to measure the amount of radiation in terms of effective number
of neutrino species $N_{\nu}^{\rm eff}$ with an error $\Delta N_{\nu}^{\rm eff} \simeq 0.25$. 
The current measured range, $N_{\nu}^{\rm eff}= 2-6$ (at $99 \%$ C.L.), seem in any case to 
confirm the presence of a thermal component of neutrinos during the early Universe stage. 

For a long time neutrinos were also thought to be the natural Dark Matter candidate. 
Massive neutrinos would indeed
yield at the present time a contribution to $\Omega_0$, given by
$\Omega_{\nu, 0}\simeq {\sum_i \, m_i /(93\,{\rm eV}\,h^2)}$,
where $m_i$ are the three neutrino masses and $h$ 
is the Hubble constant
in units of $100\,{\rm Km}\,{s}^{-1}\,{\rm Mpc}^{-1}$,
whose measurement has been recently improved by the Hubble Space Telescope
collaboration finding $h=0.742\pm 0.036$ \cite{HST}.  

Therefore, the measured value of the Dark Matter contribution to the
energy density parameter,
$\Omega_{DM, 0} = (0.227 \pm 0.014)$ \cite{WMAP7}, 
could be in principle easily explained with massive neutrinos
with $\sum_i m_i \simeq 11\,{\rm eV}$. Sufficiently large neutrino masses
could therefore solve the Dark Matter puzzle from a pure energetic point of view. 
Interestingly, neutrino oscillation experiments imply that neutrino are indeed massive
and place  a lower bound  on the sum of the neutrino masses, 
$\sum_i\, m_i \gtrsim 0.05\,{\rm eV}$.
However, from the theory of structure formation, 
we know that such light neutrinos would behave
as  `Hot Dark Matter' particles, i.e. particles that
are ultra-relativistic at the time of the matter-radiation equality
when the temperature of the Universe $T$ was $T \sim 1\,{\rm eV}$.
 
\setcounter{footnote}{4}

On the other hand current observations strongly favour the $\L$CDM model
where Dark Matter is purely `Cold', i.e. made of particles that were fully
non-relativistic at the matter-radiation equality time. They 
do not leave much room for a Hot Dark Matter component 
since even a small fraction, $\Omega_{HDM,0}/\Omega_{DM,0}\gtrsim 0.05$, 
would lead to a large-scale structure in disagreement with 
the observations.  This translates into
a very stringent limit on the sum of the neutrino masses 
\footnote{Notice that this upper bound can be easily derived from the numbers 
quoted so far: we leave it as a simple exercise for the reader.}
\cite{WMAP7} 
\be\label{WMAP7}
\sum_i\, m_i \lesssim 0.58\,{\rm eV} \hspace{10mm} (95\% \, C.L.)\, .
\ee 
Therefore, though the results from neutrino oscillation experiments could 
have represented an opportunity for neutrino masses to play  a fundamental 
role in cosmology,  with the advent of the $\L$CDM model, this became 
soon a  sort of Pyrrhic  victory
\footnote{In July 1998, when the Super-Kamiokande Japanese experiment
presented the results strongly supporting the evidence
for neutrino oscillations in atmospheric neutrinos \cite{SuperK} , the dominant cosmological
model was the so called Hot Cold Dark matter Model (HCDM), where 
both a Cold and Hot component of Dark Matter combined together to give $\O_{M,0}=\O_0= 1$
with vanishing cosmological constant. The discovery
of neutrino masses was then  announced, even in newspapers, as a confirmation
of the presence of Hot Dark Matter establishing a fundamental 
role of neutrino masses in cosmology. However, the same year,  the discovery
of the acceleration of the Universe \cite{SNIa}  marked the decline of the HCDM model
and the advent of $\L$CDM model that, as we have seen, rules 
the marginal role of a Hot Dark Matter component.
This is why the 1998 discovery of neutrino masses can be so far regarded as a 
Pyrrhic victory for the role that they play in cosmology.}.
 
However, there is another intriguing possibility 
that emerged from neutrino oscillation experiments.
The current narrow experimentally allowed window for the sum of the neutrino masses,
$0.05\,{\rm eV}\lesssim\sum_i \, m_i \lesssim 0.6\,{\rm eV}$,
seems to be optimal for these to have played a 
completely different but equally fundamental cosmological role: 
the generation of the matter-antimatter asymmetry of the Universe. 

\subsection{The matter-antimatter asymmetry of the Universe}

Before the precise observations of the CMB temperature anisotropies, 
BBN has provided a way to measure the amount of baryons in 
the Universe  during a period of time   between freeze out 
of neutron-to-proton abundance ratio and Deuterium synthesis,
corresponding approximately to $t\simeq (1-300)\,{\rm s}$. In Standard BBN,
given the value of the neutron life time, the predictions on the 
primordial nuclear abundances can be 
expressed in term of the value of the baryon-to-photon number ratio  
$\eta_{B}= (n_{B}-n_{\bar{B}})/n_{\gamma}$. 
Therefore, any measurement of a primordial nuclear abundance can
be translated, in Standard BBN, into a measurement of $\eta_{B}$.
Moreover in a Standard picture, $\eta_{B}$ is constant after the end of the BBN
and, therefore,  the value  derived from BBN is equal to the value at the present time. 
In this way $\eta_B$ can be also  related to the baryonic contribution 
to  the energy density parameter, explicitly
\be
\eta_{B} \simeq 273.6\,\Omega_{B,0}\,h^2  \times 10^{-10}  \, .
\ee
The best `baryometer' \cite{schramm} has long been known to be 
the Deuterium abundance, since this has a strong dependence
on $\eta_{B}$ ($D/H \propto \eta_B^{-1.6}$). From the measurements of the primordial
Deuterium abundance  it is found \cite{tytler}
\be
\eta_{B}^{D/H} =   (5.9 \pm 0.5) \times 10^{-10}   \, .
\ee
A completely independent measurement of $\eta_{B}$ 
during the recombination epoch coming 
from a fit of the CMB acoustic peaks and latest
results from WMAP7 data gives \cite{WMAP7}
\be\label{etaBCMB}
\eta_{B}^{CMB} = (6.2 \pm 0.15) \times 10^{-10} \, .
\ee
It should be said that, differently from the other cosmological parameters, 
it  is measured that precisely ($2.5\%$) using just CMB data. 
Moreover it is also  quite insensitive to the so called 
`priors' assumed in the $\L$CDM model (for example whether neutrinos have a mass or not)
\setcounter{footnote}{6}\footnote{The term `prior' has a statistical meaning and is more specific 
than  the generic term `assumption': it is the {\em a priori} probability distribution to 
assign to a certain variable before its measurement. An example is 
the {\em a priori} probabilities of getting head or tail when we toss a coin
without any other information,  clearly $1/2$ and $1/2$. }. 

The impressive coincidence between these two independent measurements of $\eta_B$
can be not only regarded as a great success of the hot Big Bang models
but it also strongly constrains variations   of $\eta_{B}$ between the time of BBN
$t_{BBN}={\cal O}(1s-100s)$  and the recombination time $t_{\rm rec}=
{\cal O}(10^{6}{\rm yr})$. Considering that only quite exotic effects
could have changed the value of $\eta_{B}$ until the present time 
without spoiling the
thermal equilibrium of CMB, this can therefore be regarded as a 
measurement of $\eta_{B}$ at the present time as well.


Both measurements of $\eta_B$, from Standard BBN and from CMB,
are not sensitive to the sign of $\eta_B$. Therefore, they are not
able alone to test the matter-antimatter asymmetry of the Universe. 
One could for example always  invoke, in order 
to reconcile a matter-antimatter symmetric Universe
with BBN and (more recently) CMB measurements of $\eta_{B}$,
some unspecified mechanism that segregated matter and antimatter
so that our Universe would be made of a patch of matter and antimatter domains.
However, combining the small deviations from thermal equilibrium of CMB with 
cosmic rays observations  constraining annihilations at the borders of the domains,
one can exclude such a possibility 
unless assuming very ad hoc geometrical configurations \cite{glashow}. 

Barring therefore the possibility that primordial antimatter escapes 
current observations in a very exotic way, we can conclude that $\eta_{B}$
is positive everywhere, that primordial $n_{B} \gg n_{\bar{B}}$ and that 
therefore our observable Universe is matter-antimatter asymmetric with no 
presence of primordial antimatter whatsoever. 

At first sight this observation does not seem to be a real issue since one could 
after all  assume that the observed value of $\eta_{B}$ is just an initial set up value not 
requiring any explanation. However, this legitimate minimal solution is not compatible
with the $\L$CDM model where an inflationary stage is a crucial ingredient. 
A qualitative argument is that within inflation our observable Universe would correspond
to the magnification of such a tiny region, that thus can be assumed to be simply empty
at the end of Inflation, except for the presence of a vacuum energy density
responsible for the Inflationary stage itself. Therefore, according to Inflation,
all the matter and the antimatter that we observe today,  including the matter-antimatter 
asymmetry, had to be  produced after Inflation.

In this way, one arrives at the conclusion 
that the baryon asymmetry must have been dynamically generated either 
after inflation and prior to the BBN epoch or in any case during the 
latest period of the inflationary stage, i.e. to the necessity of 
a model of baryogenesis in complete agreement with the much earlier Sakharov's idea. 

In his famous prescient paper \cite{sakharov}, Sakharov identified, either explicitly or
implicitly, three famous necessary conditions, bearing now his name, for
a successful model of baryogenesis: (i) the existence of an elementary process that violates
the baryon number,  (ii) violation of charge conjugation $C$ and of $CP$, where $P$ indicates parity transformation, 
(iii) a departure from thermal equilibrium during baryogenesis. Notice that the departure from thermal
equilibrium has to be permanent, since otherwise the baryon asymmetry
would be subsequently washed-out if thermal equilibrium is restored.


It was realised in the mid 80's that, within the SM, 
the Sakharov conditions are all fulfilled to some level. It became then  
important to give a quantitative estimation, finding the 
most stringent conditions for a baryon asymmetry to be 
generated at the level of the observed value.
The most efficient model is realised by electroweak baryogenesis \cite{EWB}.
A strong departure from thermal equilibrium is obtained if at the
electroweak symmetry breaking  a strong first order phase transition
occurs accompanied by the nucleation of bubbles. Inside the expanding  bubbles 
electroweak symmetry is broken, outside is unbroken. Particles and anti-particles
outside the bubble cross the bubble wall with a different rate in a way that an asymmetry
is generated inside the bubble. If the transition is strong enough such an asymmetry
is not washed-out and can survive until the present time. Unfortunately the condition
for a strong first order phase transition implies a stringent upper bound on the 
Higgs mass $m_H \lesssim 40\,{\rm GeV}$ clearly at odds with the LEP lower bound
$m_H \gtrsim 114\,{\rm GeV}$. 

Nobody has found another model of baryogenesis not requiring
some extension of the SM. For this reason the matter-antimatter
asymmetry is nowadays regarded as strong evidence of new physics. 
\setcounter{footnote}{7}\footnote{For example, an electroweak baryogenesis model can be made successful within a popular extension 
of the SM, the so called minimal supersymmetric standard model (MSSM). 
In this case a  conservative upper bound on the (light) Higgs mass,
$m_H\lesssim 127\,{\rm GeV}$, is found \cite{nardini} but
values $120\,{\rm GeV} \lesssim M_H \lesssim 127 \,{\rm GeV}$
are possible only within  quite a small corner in the parameter space.
For this reason the recent data from the Large Hadron Collider, 
favouring a SM-like Higgs boson mass around $125\,{\rm GeV}$, 
are very challenging for electroweak baryogenesis within the MSSM.
It is important to notice this point because electro-weak baryogenesis is 
probably the strongest competitor of leptogenesis .}

\section{Neutrino masses and mixing}

The discovery of neutrino masses and mixing in neutrino oscillation
experiments is the first clear indication of new physics in laboratory experiments.
This has, therefore,  greatly raised interest in a model of baryogenesis 
relying  on the most  popular way to extend the SM in order to explain
why neutrinos are massive but at the same time much lighter than all
other fermions: the see-saw mechanism. In this section we first discuss
the main experimental results on neutrino masses and mixing parameters
and then we briefly discuss the see-saw mechanism. This will be crucial
to understand the main features and also the importance of leptogenesis 
within modern particle physics. 

\subsection{Experiments}

Neutrino oscillation experiments have established two fundamental 
properties of neutrinos \cite{king}. The first one is that  neutrino flavour eigenstates
$\nu_{\alpha}$ ($\alpha=e,\mu,\tau$) 
do not coincide with  neutrino mass eigenstates $\nu_i$ ($i=1,2,3$) but
are obtained applying to these a unitary transformation described by a 
$(3\times 3)$ unitary leptonic mixing matrix $U$,
\be\label{mixing}
\nu_{\alpha} = \sum_{i} \, U_{\alpha i}\,\nu_i \, .
\ee
 The leptonic mixing
matrix\setcounter{footnote}{8}\footnote{It is also often called the Pontecorvo-Maki-Nakagata-Sakata matrix and 
indicated with the symbol $U_{PMNS}$.}
 is usually parameterised in terms of  6 physical parameters, 
three mixing angles, $\theta_{12}$,  $\theta_{13}$ and $\theta_{23}$, and three phases,
the two Majorana phases, $\rho$ and $\sigma$, and the Dirac phase $\delta$,
\be\label{Umatrix}
U=\left( \begin{array}{ccc}
c_{12}\,c_{13} & s_{12}\,c_{13} & s_{13}\,e^{-{\rm i}\,\d} \\
-s_{12}\,c_{23}-c_{12}\,s_{23}\,s_{13}\,e^{{\rm i}\,\d} &
c_{12}\,c_{23}-s_{12}\,s_{23}\,s_{13}\,e^{{\rm i}\,\d} & s_{23}\,c_{13} \\
s_{12}\,s_{23}-c_{12}\,c_{23}\,s_{13}\,e^{{\rm i}\,\d}
& -c_{12}\,s_{23}-s_{12}\,c_{23}\,s_{13}\,e^{{\rm i}\,\d}  &
c_{23}\,c_{13}
\end{array}\right)
\cdot {\rm diag}\left(e^{i\,\rho}, 1, e^{i\,\sigma}
\right)\, ,
\ee
where $s_{13} \equiv \sin\theta_{13}$ and $c_{13}\equiv\cos\theta_{13}$.
It should be noticed that the three phases are $C\!P$ violating if they are  
not integer multiples of $\pi$.
The second important property established by neutrino oscillation experiments
is that neutrinos are massive. More particularly, defining the three neutrino masses
in a way that $m_1 \leq m_2 \leq m_3$, the neutrino oscillation experiments 
measure two mass squared differences that we can indicate with
$\Delta m^2_{\rm atm}$ and $\Delta m^2_{\rm sol}$ since historically the first
one has been first measured in atmospheric neutrino experiments and the second one in
solar neutrino experiments. 

There are two options that are both allowed by
current experiments. A first option is the so called `normal ordering' (NO) and in this case
one has
\be
m_3^{\, 2} - m^{\, 2}_2 = \Delta m^2_{\rm atm}  \hspace{5mm} \mbox{\rm and} 
\hspace{5mm}
m_3^{\, 2} - m^{\, 2}_1 = \Delta m^2_{\rm  sol} \, ,
\ee
while a second option is represented by so called `inverted ordering' (IO) and in this case
\be
m_2^{\, 3} - m^{\, 2}_2 = \Delta m^2_{\rm sol}  \hspace{5mm} \mbox{\rm and} \hspace{5mm}
m_2^{\, 2} - m^{\, 2}_1 = \Delta m^2_{\rm  atm} \,  .
\ee
It is convenient to introduce the atmospheric neutrino mass scale
$m_{\rm atm} \equiv \sqrt{\Delta m^2_{\rm atm}+\Delta m^2_{\rm  sol}}
= (0.049\pm 0.001)\,{\rm eV}$ and the solar neutrino mass scale 
$m_{\rm sol} \equiv \sqrt{\Delta m^2_{\rm  sol}}=(0.0087 \pm 0.0001)\,{\rm eV}$ 
\cite{oscillations}.

The measurements of $m_{\rm atm}$ and $m_{\rm sol}$ 
are not sufficient to fix all three neutrino masses. 
If we express them in terms of the lightest neutrino 
mass $m_1$ we can see from Fig.~1 that while 
 $m_2 \geq m_{\rm sol}$ and $m_3 \geq m_{\rm atm}$, the lightest
neutrino mass can be arbitrarily small implying that the lightest neutrino
could  still be even massless. 

The lower bounds for $m_2$ and $m_3$ 
are saturated when $m_1 \ll m_{\rm sol}$. In this case one has hierarchical
neutrino models, either normally ordered, and in this case $m_2 \simeq m_{\rm sol}$
and $m_3\simeq m_{\rm atm}$, or inversely ordered, and in this case 
$m_2\simeq \sqrt {m^2_{\rm atm}-m^2_{\rm sol}} \simeq m_3\simeq m_{\rm atm}$.  
On the other hand for $m_1\gg m_{\rm atm}$ one obtains the limit of quasi-degenerate 
neutrinos when all three neutrino masses can be arbitrarily close to each other. 

The lightest neutrino mass $m_1$ is however upper bounded by absolute neutrino mass scale
experiments. Tritium beta decays experiments \cite{tritium} place an upper bound on the effective electron  neutrino mass $m_{\nu_e} \lesssim 2\,{\rm eV}$ ($95 \%$ C.L.) that translates
into the same upper bound on $m_1$.  This is derived from model independent
kinematic considerations that apply independently whether neutrinos have a Dirac or 
Majorana nature. 

Neutrinoless double beta decay ($0\nu\b\b$) experiments 
place a more stringent upper bound on the effective $0\nu\b\b$ Majorana neutrino mass 
$m_{ee}\lesssim (0.34-0.78)\,{\rm eV}$\,($95\% C.L.$)  \cite{bb0n}
that also translates into the same upper bound  on $m_1$. 
Here the wide range is due to  theoretical uncertainties in the calculations
of the involved nuclear matrix elements.  This upper bound applies
only if neutrinos are of Majorana nature, that is however a prediction of 
the see-saw mechanism and, therefore, it will be relevant to our discussion. 
\begin{figure}
\begin{center}
\begin{minipage}{150mm}
\begin{center}
\subfigure[\empty]{
\resizebox*{10cm}{!}{\includegraphics{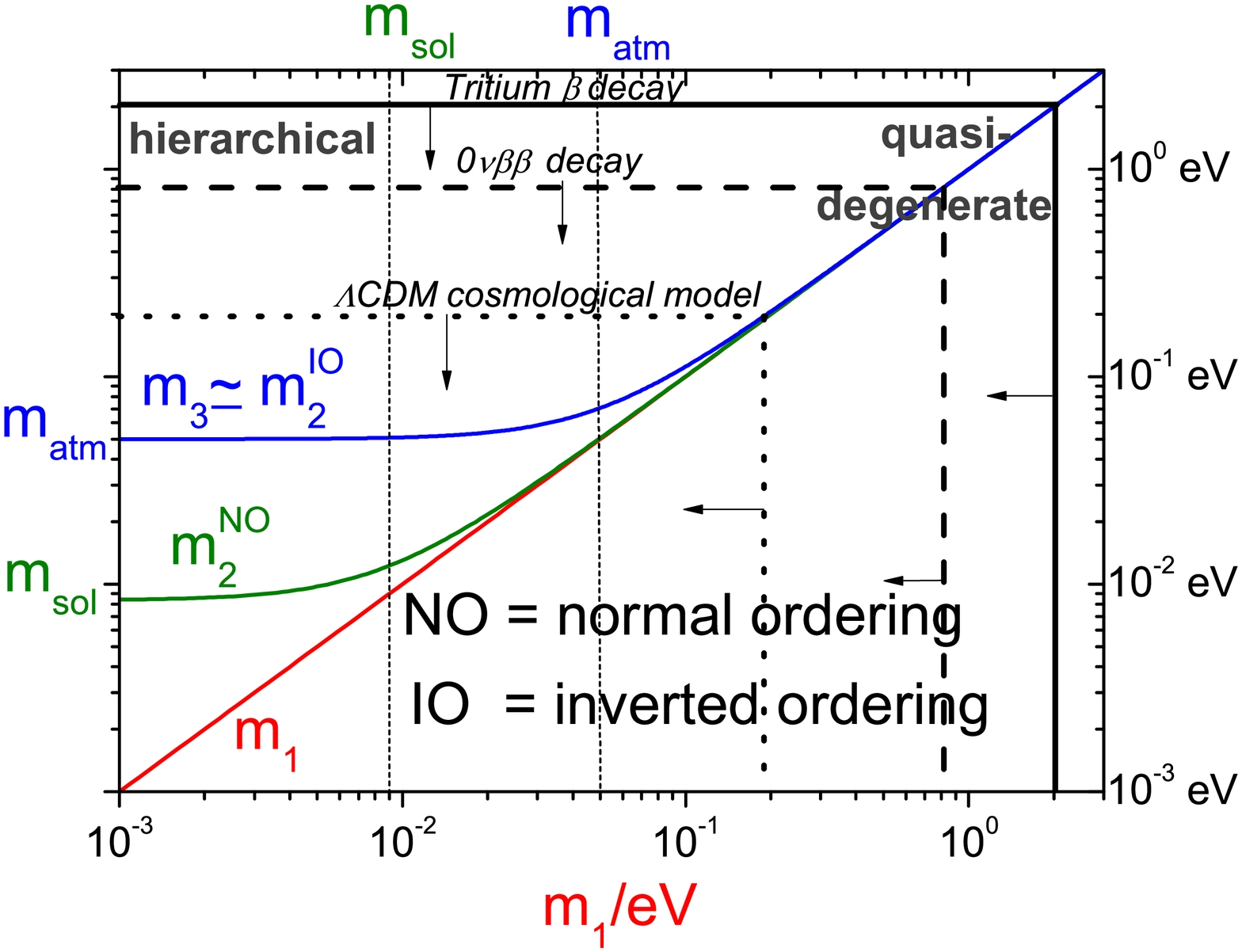}}}
\caption{Neutrino masses $m_i$ versus the lightest neutrino mass $m_1$.
The three upper bounds (at $95\% C.L.$)  discussed in the body text
from absolute neutrino mass scales phenomenologies are also indicated.}
\end{center}
\end{minipage}          
\end{center}
\end{figure}
As we discussed, from cosmological observations, 
within the $\L$CDM model, one obtains a very stringent upper bound on the 
sum of the neutrino masses eq.~(\ref{WMAP7}) that translates into the
most stringent upper bound that we currently have on the 
lightest neutrino mass $m_1 \lesssim 0.19\,{\rm eV} (95 \% C.L.)$, 
almost excluding quasi-degenerate neutrino models. 

 \subsection{Theory: the see-saw mechanism}

A minimal extension of the SM able to explain not only why neutrino
are massive but also why they are much lighter than all the other fermions,
is given by the see-saw mechanism \cite{see-saw}.  
The SM is a quantum field theory \cite{mandlshaw}.  All physical laws can ultimately be extracted by
a fundamental object that is the Lagrangian, exactly like in classical systems
\setcounter{footnote}{9}\footnote{The SM Lagrangian is  such a fundamental object in modern particle physics
that it is even a source of merchandising, and can be found printed on CERN T-shirts.}. 
The difference is that while in classical
systems the fundamental entities are either point-like massive particles or classical fields (e.g. the electromagnetic field), 
in a quantum field theory there is only one kind of fundamental entities, the quantum fields.
These are operators acting on the quantum states. The quantum states that correspond 
to the excitations of the quantum fields correspond to 
the free elementary particles.  In this way both matter
and radiation are described  in the same way. 

The Lagrangian contains not only (kinetic) terms describing the evolution
of the free fields but also  terms describing interactions among fields. 
These terms are not arbitrary but emerge once particular
`symmetries' are imposed on the Lagrangian. In modern quantum field theories the symmetries are
elements of a `gauge group' and the SM is based on symmetries of a non-abelian gauge group
called $SU(3)\times SU(2) \times U(1)$.  Within the SM, neutrinos play quite
a special role  since they are
massless and since they can have only one helicity, the left-handed one
\setcounter{footnote}{10}\footnote{Actually it can be proven that the two things are related: since there are no RH
neutrinos in the SM, then the ordinary left-handed neutrinos we observe have to be exactly massless in the SM.}.
This means that their
spin vector always points opposite to the direction of motion (while for the anti-neutrinos is the
other way around). Therefore, in the SM lagrangian there are no right-handed (RH) neutrinos (and no
left-handed (LH) anti-neutrinos).

In a minimal version of the see-saw mechanism one adds $N$
RH neutrinos $N_{iR}$ to the SM lagrangian $\mathcal{L}_{\rm SM}$ . 
These RH
neutrinos can interact with the SM leptons
\setcounter{footnote}{11}\footnote{The SM leptons are given by  the three charged leptons, electrons, muons and tauons, plus the three species of
neutrinos, respectively one for each charged lepton in a way to have
three lepton doublets.}  through so called  Yukawa interactions described by a complex matrix $h$ called the
Yukawa coupling matrix.  At the same time  one can also add a so called 
right-right Majorana mass term that violates lepton number and that is described by
another complex matrix, the Majorana mass matrix $M$. 
In this way the see-saw lagrangian can be written as
\be\label{lagrangian}
\mathcal{L}  =  \mathcal{L}_{\rm SM} 
+i \sum_{i}\overline{N_{i R}}\g_{\m}\partial^{\m} N_{i R} -
\sum_{\a, i}\overline{\ell_{\a L}}\, h_{\a i}\, N_{i R} \,{\f} - 
 {1\over 2}\, \sum_{i,j} \overline{N_{i R}^c}\, M_{ij} \, N_{j R} +h.c. \, ,
\ee
where $\a=e,\mu,\tau$ and $i,j=1,\dots,N$ and `h.c.' denotes hermitian conjugated terms. 
In this equation the ${\ell}_{\a L}$'s denote the fields associated to the 
so called lepton doublets (one for each flavour $\a$), each including one
charged lepton $\alpha$ (electrons, muons and tauons) and one neutrino $\nu_{\a}$,  
while the symbol $\f$ denotes the Higgs field.  

We do not know in general the number $N$ of RH neutrinos. We just know 
that, in order to reproduce correctly the observed neutrino masses and mixing parameters
with the see-saw mechanism that we are going to discuss in a moment, 
we need at least two RH neutrinos (i.e. $N\geq 2$). 
For definiteness we will consider the case $N=3$.
This is most attractive case corresponding to have one RH neutrino for each generation 
in the SM, as, for example, nicely predicted by $SO(10)$ grand-unified  models
\footnote{Grand-unified models 
are models beyond the SM where at a some very high
energy grand-unified scale $M_{GUT}$ the three fundamental interactions of the SM,
the electromagnetic, weak and strong interactions, unify and are described in terms of the
same gauge group, e.g. $SO(10)$ for $SO(10)$ models. In typical models the grand-unified scale 
is $M_{GUT} \sim 10^{16}\,{\rm GeV}$.}.
Notice, however, that all current
data from low energy neutrino experiments are also consistent
with a more minimal two RH neutrino model, as we will discuss in more detail.

The addition of RH neutrinos  in the Lagrangian 
with Yukawa couplings $h$ to leptons, follows exactly 
the same recipe that allows all massive fermions in the SM 
to acquire a mass through the so called Higgs mechanism. 
Below a certain critical temperature the neutral Higgs field acquires 
a non zero vacuum expectation value $v$ that breaks the electroweak
symmetry of the theory and acts as a background Higgs field that fills
the whole space. This is analogous to what happens in ferromagnets
where below a certain critical temperature there is a non-vanishing magnetic
field that breaks rotational symmetry.  This
  spontaneous electroweak symmetry breaking results in fermion masses
  due to their interaction with the Higgs background field $v$. The mechanism
  requires the existence of RH particles. Therefore, in the SM,
  neutrinos do not acquire any mass simply because they do not couple to $v$. 
  However, when RH neutrinos are added with Yukawa couplings $h$
to $v$,  a neutrino Dirac mass term $m_D=v\,h$
is generated as for all the other massive fermions.  If this were the whole story, 
one could easily understand why neutrinos are massive but not why
they are orders-of-magnitude lighter than all the other particles. 
However, there is another element to be considered. Differently from all the other
massive fermions,  neutrinos are neutral and in general the Majorana mass 
term in the Lagrangian eq.~(\ref{lagrangian}) is also present.  In this way
the neutrino masses that we measure would be the result of an interplay between 
the usual Dirac neutrino mass term and the  Majorana neutrino mass term.  

Indeed one can show that in the so called `see-saw limit', for $M\gg m_D$
\setcounter{footnote}{13}\footnote{The condition $M\gg m_D$ can appear ambiguous if not totally meaningless since in general $M$ and $m_D$
are complex matrices. However, it is a compact conventional notation to indicate
that all eigenvalues of $M$ have to be much
greater than all eigenvalues of $m_D$, or, equivalently, that the smallest 
eigenvalue of $M$ has to be much greater than the largest eigenvalue of $m_D$.}, 
the spectrum
of (ordinary) neutrino masses splits into a light set given by the eigenvalues $m_1<m_2<m_3$
of the neutrino mass matrix
\be\label{see-saw}
m_{\nu} = - m_D\,{1\over M}\,m_D^T \,
\ee
and into a  (new) heavy set coinciding with very good
approximation with the eigenvalues $M_1 < M_2 < M_3$ of the Majorana mass matrix $M$
and corresponding to eigenstates $N_i\simeq N_{i R}$. Notice that in this formula
${1\over M} \equiv M^{-1}$. This notation emphasises the pictorial `see-saw' analogy (see Fig.~2). 
\begin{figure}
\begin{center}
\begin{minipage}{150mm}
\begin{center}
\subfigure[\empty]{
\resizebox*{10cm}{!}{\includegraphics{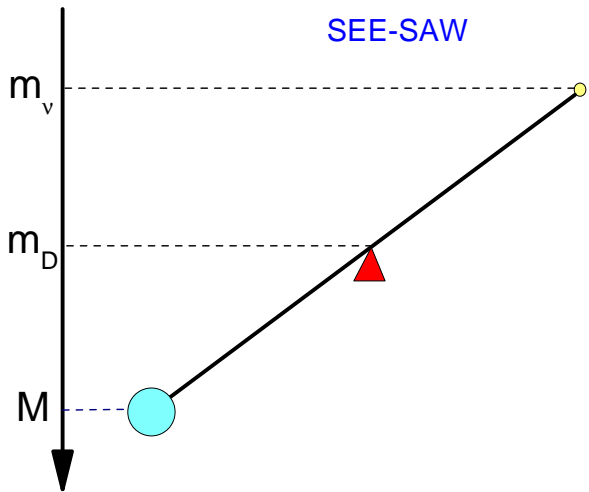}}}
\caption{A pictorial representation of the see-saw formula eq.~(\ref{see-saw}). The large Majorana
mass scale $M$, related to some new physics scale like for example the grand-unified scale $M_{GUT}$, 
pushes up the
light neutrino mass scale pivoting on $m_D$.}
\end{center}
\end{minipage}          
\end{center}
\end{figure}
 Consider indeed a toy model case where there is only one neutrino mass scale
(it can be regarded as an approximation limit case for $m_{\rm sol}/m_{\rm atm}\rightarrow 0$)
and all matrices reduces to simple numbers. If one chooses for the neutrino Dirac mass
a value $m_D \sim v \sim 100\,{\rm GeV}$, the electroweak scale dictating the masses
of all other fermions, then one can see that the atmospheric neutrino mass scale $m_{\rm atm}$
is reproduced for a  value of $M\sim 10^{14-15}\,{\rm GeV}$, intriguingly very close (order-of-magnitude wise) to the
grand-unified scale $M_{GUT}\sim 10^{15-16}\,{\rm GeV}$: as in a see-saw, the huge RH neutrino mass $M$, 
`sitting downstairs' in the see-saw formula (\ref{see-saw}) and pivoting on $m_D$, 
is able to push down the light neutrino mass $m_{\nu}$. 

It should then be noticed a very important implication of the see-saw mechanism: the light neutrino mass scale
is not a fundamental scale, but an algebraic by-product of the combination of two much bigger scales.
Within this picture, neutrino physics becomes then a portal, at low energies, 
of new physics at very high energies.

If we now consider the realistic case with three neutrino masses,
the $(3\times 3)$ symmetric neutrino mass matrix $m_{\nu}$ is diagonalised by a unitary matrix $U$, explicitly
\be
D_m \equiv {\rm diag}(m_1,m_2,m_3) = -U_{\nu}^{\dagger}\,m_{\nu}\,U_{\nu}^{\star} \, .
\ee
For all practical purposes,  the matrix $U_{\nu}$ here 
can be assumed to coincide with the leptonic mixing 
matrix $U$ defined in the eq.(\ref{mixing}). 
 The masses of the RH neutrinos
are very model dependent. However it is fair to say that in order to
reproduce the measured light neutrino mass scales  and 
barring fine tuned choices of the parameters, typical proposed 
models give an heaviest RH neutrino mass $M_3\gtrsim 10^{14}\,{\rm GeV}$. 
This is the minimal version of the see-saw mechanism,
often indicated as type I see-saw mechanism. 

The see-saw formula eq.~(\ref{see-saw}) can be recast as an orthogonality  
condition for a matrix $\Omega$ that, in a basis where simultaneously the 
charged lepton mass matrix and the Majorana mass matrix are diagonal,
provides a useful parametrisation of the neutrino Dirac mass matrix \cite{casas},
\be\label{orthogonal}
m_D=U\,D_m^{1/2}\,\O\, D_M^{1/2} \,  ,
\ee
where we defined $D_m\equiv {\rm diag}(m_1,m_2,m_3)$ and
$D_M\equiv {\rm diag}(M_1,M_2,M_3)$.  The orthogonal matrix $\O$ encodes 
the properties of the heavy RH neutrinos, the three life times and the
three total $C\!P$ asymmetries.  This parametrisation is quite useful since,
given a model that specifies $m_D$ and the three RH neutrino masses $M_i$, 
through it one can easily  impose both the experimental
information on the $9$ low energy neutrino parameters in $U$ and $m_i$ and, as we will
see, also the requirement of successful leptogenesis. 

\section{The simplest scenario of leptogenesis: vanilla leptogenesis}

In the most general approach the asymmetry depends on all the 18 see-saw parameters that
can be identified with nine low energy parameters (three light neutrino masses plus 
the six parameters in the leptonic mixing matrix)
and  nine high energy neutrino parameters associated to the 
three RH neutrino properties (masses, lifetimes and total $C\!P$ asymmetries). 
Moreover the calculation itself presents different technical difficulties.
However, there is  a simplified  scenario \cite{fy,upperbound,bounds} that grasps 
most of the main features of leptogenesis and is able to highlight  important connections
with the low energy neutrino parameters in an approximated way. 
We will refer to this  scenario as  `vanilla leptogenesis'. 
We discuss  the main features, assumptions and approximations. 
The discussion can be conveniently split into two parts: in the first
part we discuss how the  abundance of the RH neutrinos is calculated during the expansion
of the Universe, while in the second part we describe how the baryon asymmetry is calculated. 

\subsection{Assumptions of minimal  leptogenesis}

Leptogenesis  belongs to a class of models of baryogenesis where the asymmetry is generated from  the 
out-of-equilibrium decays of very heavy particles, quite interestingly the same class as the first model
proposed by Sakharov belongs. This class of models became very popular with the 
advent of grand-unified theories (GUTs) that provided a specific well definite and motivated framework.
In GUT baryogenesis models the very heavy
particles are the same new gauge bosons predicted by GUTs.
However,  the final asymmetry depends on too many untestable parameters, 
so that imposing successful baryogenesis does not lead to 
compelling experimental predictions.
This lack of predictability is made even stronger 
considering that the decaying particles are 
too heavy to be produced thermally and one has therefore to invoke
a non-thermal production mechanism of the gauge bosons. 
This is because while the mass of the gauge bosons is 
about the grandunification scale, $M_X \sim 10^{15-16}\,{\rm GeV}$,  
the reheating temperature at the end of inflation $T_{RH}$ cannot be higher
than $\sim 10^{15}\,{\rm GeV}$ from CMB observations. 
The reheating temperature is  the initial value of the temperature at the beginning of  
the radiation dominated regime after Inflation. Below this temperature 
the inflationary stage can, therefore, be considered  concluded.  

The minimal (and original) version of leptogenesis is based on the 
type I see-saw mechanism and the  asymmetry is produced 
by  the three heavy RH neutrinos. 
We will call `minimal leptogenesis scenarios' those scenarios
where a type I see-saw mechanism and a thermal production
of the RH neutrinos (thermal leptogenesis), implying that $T_{RH}$ is comparable at least to the lightest
RH neutrino mass $M_1$, are assumed. At these high temperatures the RH neutrinos can be produced
by the Yukawa interactions of leptons and Higgs bosons in the thermal bath.
Since in most models of neutrino masses embedding the type I see-saw mechanism
the lightest RH neutrino mass $M_1 \ll 10^{15}\,{\rm GeV}$, the condition
of thermal leptogenesis can, therefore, be satisfied compatibly with the 
above-mentioned upper bound, $T_{RH}\lesssim 10^{15}\,{\rm GeV}$,
from CMB observations. 

After their production, the RH neutrinos decay either 
into leptons $N_i \ra {\ell}_i + \phi^{\dagger}$, with a decay rate $\Gamma_i$, 
or into anti-leptons and Higgs bosons $N_i \rightarrow \bar{\ell}_i + \phi$,
with a decay rate $\bar{\Gamma}_i$. Notice that here the leptons ${\ell}_i$
and the anti-leptons $\bar{\ell}_i$  are  {\em by definition} the
leptons and the anti-leptons produced by the decays of the RH neutrinos. 
In general, as we will discuss in detail later on, they are described by quantum states that 
are a linear  combination of the flavour eigenstates describing the leptons ${\ell}_{\a}$ ($\alpha= e,\mu,\tau$) 
associated to the left-handed lepton doublet fields ${\ell}_{\a L}$ written  in the lagrangian eq.~(\ref{lagrangian}).
Since both the RH neutrinos $N_i$ and the Higgs bosons $\phi$ 
do not carry lepton number, both inverse processes and decays 
violate lepton number ($|\Delta L|=1$)  and in general $C\!P$ as well. 
It is also important to notice that they also violate the $B-L$
number, ($|\D(B-L)|=1$).
At temperatures $T\gg 100\,{\rm GeV}$  non perturbative SM processes 
called sphalerons are in equilibrium. They violate both lepton and baryon number while 
they still conserve the $B-L$ asymmetry.
In this way the lepton asymmetry produced in the elementary processes is 
reprocessed in a way that at the end approximately $1/3$ of the $B-L$ asymmetry 
is in the form of a baryon asymmetry and $-2/3$ of the $B-L$ asymmetry
is in the form of a lepton number:  two Sakharov conditions 
are, therefore, clearly satisfied. The third Sakharov condition, the departure from thermal equilibrium,
is also satisfied since some fraction of the decays occurs out-of-equilibrium. In this way
part of the generated asymmetry survives to the wash-out from inverse processes.

\subsection{Calculation of the RH neutrino abundances}

In the calculation of the RH neutrino abundances a key 
role is played by the decay parameters $K_i$. 
They are defined as the ratio of the total decay width of the RH neutrinos $N_i$,
$\widetilde{\Gamma}_i\equiv (\Gamma_i + \bar{\Gamma}_i)_{T\ll M_i}$, to the
(Hubble) expansion rate of the Universe $H$  at $T=M_i$, when the $N_i$'s
start to become non-relativistic 
\setcounter{footnote}{14}\footnote{During the early Universe expansion the temperature drops down 
approximately as $T \propto 1/t^2$, where $t$ is the time elapsed from the
Big Bang.}, explicitly
\setcounter{footnote}{15}\footnote{The RH neutrino decay widths can be expressed in terms of the Yukawa
coupling matrix as $\widetilde{\Gamma}_i  = M_i\,(h^{\dagger}\,h)_{ii}/(8\pi)$.}
\be
K_i \equiv   {\widetilde{\Gamma}_i \over H(T=M_i)} \, .
\ee
Considering that the age of the Universe is given by $t=H^{-1}/2$, while the life-times
of the RH neutrinos, by definition, are given by $\tau_i = \widetilde{\Gamma}_i^{-1}$, the decay parameters
can also be regarded as a ratio between the age of the Universe when they become non-relativistic, $t(T=M_i)$,
and the $\tau_i$'s. Therefore, for $K_i \gg 1$,
the RH neutrinos life-time is much shorter than $t(T=M_i)$
and they (on average) decay and inverse-decay many times before they
become non-relativistic. In this situation the RH neutrino abundance tracks 
very closely the equilibrium distribution. On the other hand, for $K_i \ll 1$,
the bulk of the RH neutrinos will decay completely out-of-equilibrium, 
when they are already fully non-relativistic
and their equilibrium abundance is exponentially suppressed by the Boltzmann factor. 

A quantitative calculation of the RH neutrino abundances  (and also of the baryon asymmetry
as we will discuss in the next subsection) requires a kinetic description. 
In the `vanilla scenario' this is done with the use of
Boltzmann equations averaged over momenta (often called `rate equations'). 
Let us indicate with $N_{N_i}$ the RH neutrino abundances
and with $N_{N_i}^{\rm eq}$ their thermal equilibrium values. 
The Boltzmann equations for the $N_{N_i}$'s  are then  given by
\be\label{BERHNabundances}
{dN_{N_i}\over dz}  =  -D_i\,(N_{N_i}-N_{N_i}^{\rm eq}) \,  ,
\ee
where $z \equiv M_1/T$. Notice that we are calculating the abundance $N_X$ of 
any particle number or asymmetry $X$ in a portion of co-moving
volume containing one heavy neutrino in ultra-relativistic thermal equilibrium,
so that $N^{\rm eq}_{N_i}(T\gg M_i)=1$.

Inroducing $x_i\equiv M_i^2/M_1^2$, the decay factors $D_i$ are defined as
\be
D_i \equiv {\G_{{\rm D},i}\over H\,z}=K_i\,x_i\,z\,
\left\langle {1\over\gamma_i} \right\rangle   \, ,
\ee
where $H$ is the expansion rate. The total decay rates,
$\G_{{\rm D},i} \equiv \G_i+\bar{\G}_i$,
are the product of the decay widths $\widetilde{\Gamma}_i$ times  the
thermally averaged dilation factors
$\langle 1/\gamma\rangle$, given by the ratio
${\cal K}_1(z)/ {\cal K}_2(z)$ of the modified Bessel functions.
It should be noticed how for each RH neutrino species $N_i$, the 
associated decay parameter $K_i$ is the only parameter that determines
its abundance $N_{N_i}$ in addition to  the initial value $N_{N_i}^{\rm in}$. 

In Fig.~3 we show the result for the lightest RH neutrino abundance $N_{N_1}$
\setcounter{footnote}{16}\footnote{It will be clear in the next subsection why we focus on the lightest
RH neutrino $N_1$.} and  for the two cases $K_1 = 10^{-2} \ll 1 $ and
$K_1 = 10 \gg 1$.  The initial RH neutrino abundance $N_{N_1}^{\rm in}$, typically at some $z_{\rm in}\ll 1$, 
is in general an unknown parameters and it has been here chosen equal to
the thermal equilibrium value. 
It can be clearly seen how in the first case
the equilibrium abundance closely tracks the equilibrium value, while
in the second case there is a clear departure from thermal equilibrium.
\begin{figure}
\begin{center}
\begin{minipage}{150mm}
\begin{center}
\subfigure{\resizebox*{72mm}{!}{\includegraphics{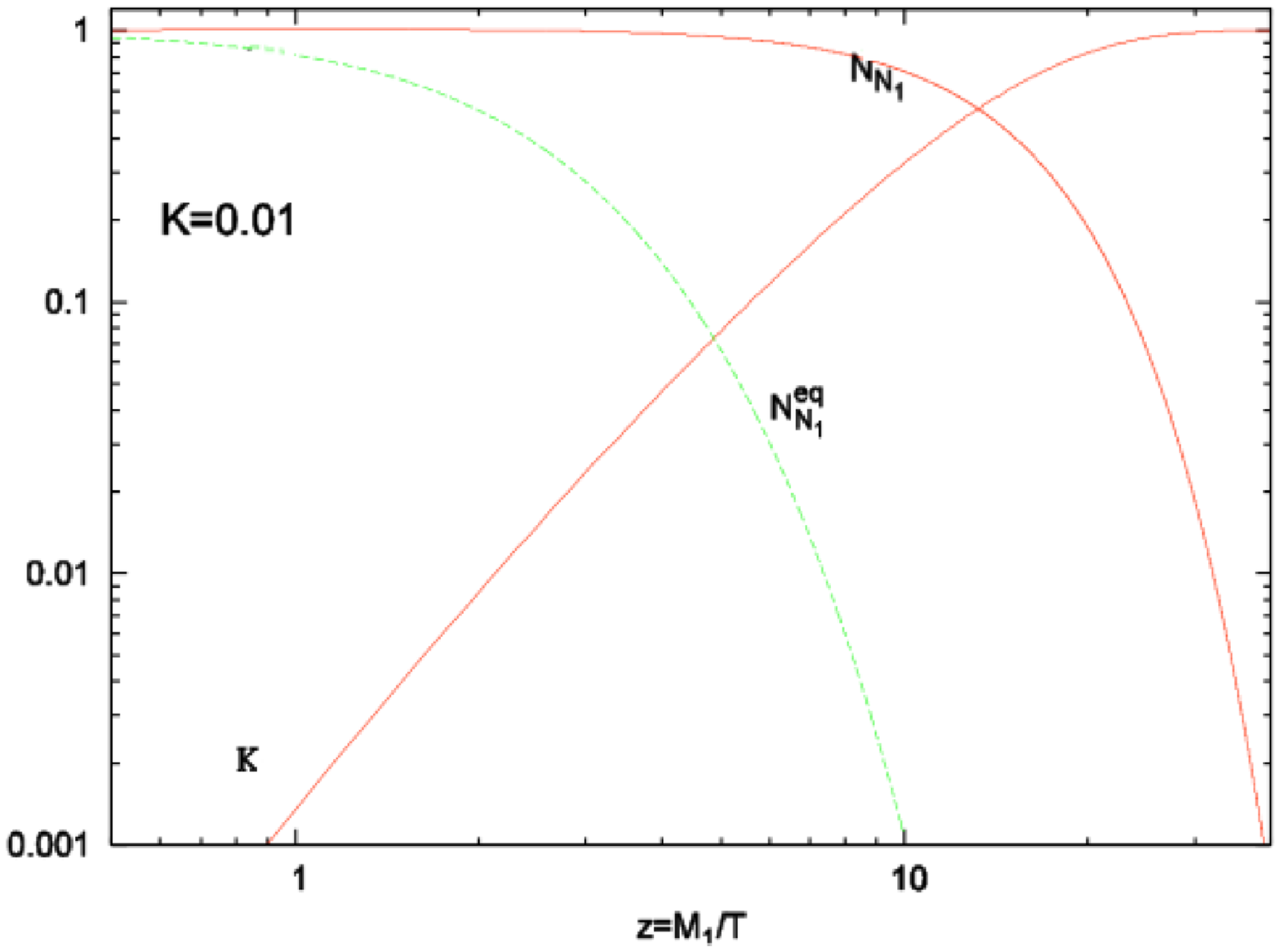}}} \hspace*{2mm}
\subfigure{\resizebox*{72mm}{!}{\includegraphics{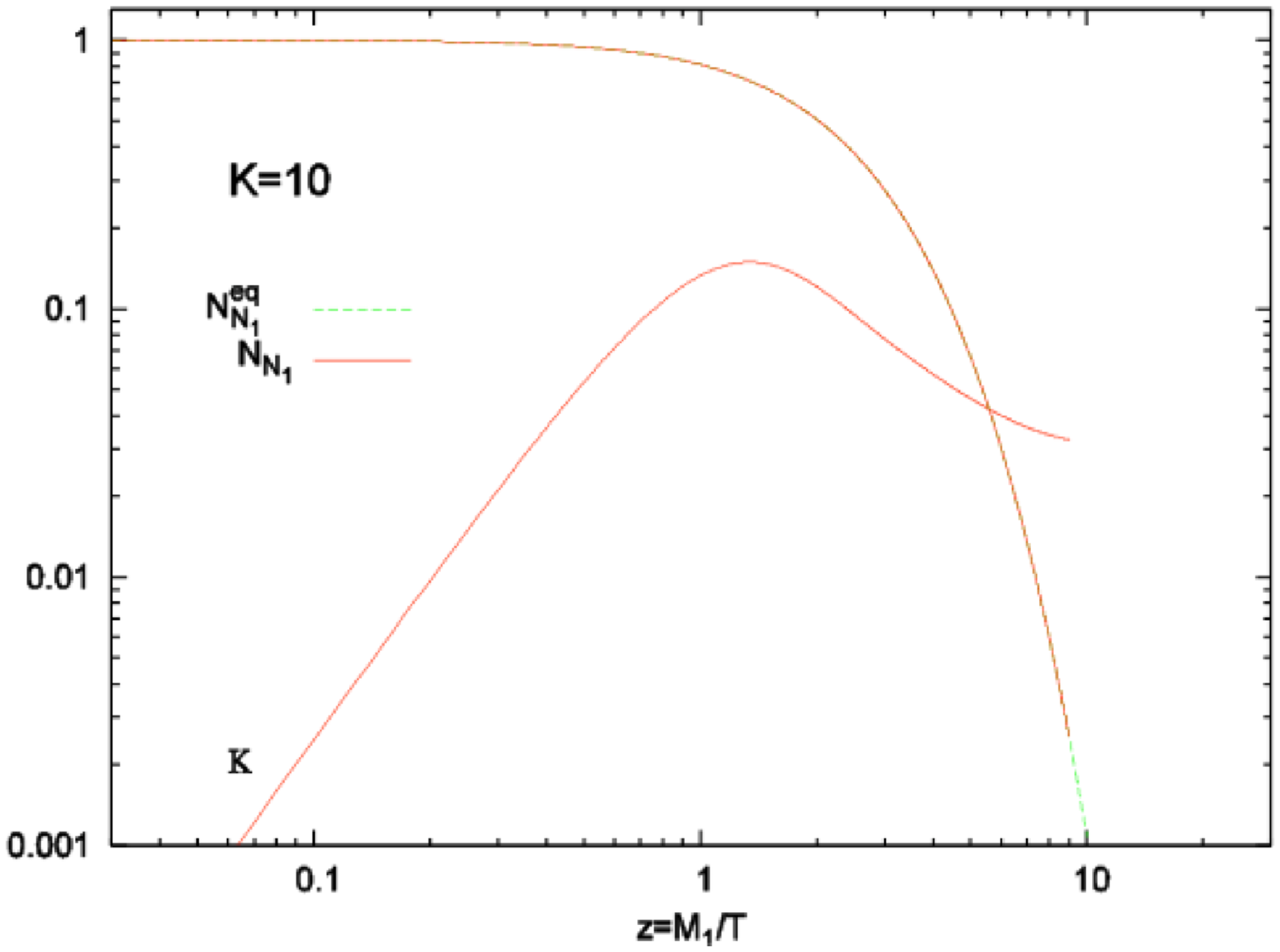}}}
\vspace*{-25mm}
\caption{Evolution of the RH neutrino abundance $N_{N_1}$ (red line) and of 
the efficiency factor $\kappa$ (red line) with the temperature for two different
values of the decay parameter $K_1$: (left) $K_1 = 0.01$ and (right) $K_1 = 10$.
The green dotted line is the thermal equilibrium abundance $N_{N_1}^{\rm eq}$: in the right panel
it is hardly visible since it is closely tracked by $N_{N_1}$.}
\end{center}
\end{minipage}          
\end{center}
\end{figure}

\subsection{Calculation of the baryon asymmetry}

Let us now turn to discuss how the baryon asymmetry is calculated in the vanilla scenario. 
The flavour composition of the leptons ${\ell}_i$ and anti-leptons $\bar{\ell}_i$
produced by (or producing) the RH neutrinos is assumed to have no influence on the
final value of the asymmetry and is therefore neglected 
(assumption of `unflavoured' or `one-flavoured' leptogenesis). 
This is equivalent to say that the only relevant quantity is the total lepton asymmetry, 
i.e. the difference between the
total number of leptons and the total number of anti-leptons irrespectively
whether these leptons are electron, muon or tauon doublets (at these
temperatures the electroweak symmetry is not broken). 
As we said at very high temperatures $T\gg 100\,{\rm GeV}$,
part of the total lepton asymmetry is rapidly converted into a baryon asymmetry
by sphaleron processes. However sphaleron processes conserve  the $B-L$ asymmetry
and it can be shown that  at the end approximately the final baryon asymmetry
$N_B^{\rm f} \simeq N_{B-L}^{\rm f}/3$
while the final lepton asymmetry   $N^{\rm f}_L \simeq -2\,N_{B-L}^{\rm f}/3$. 

For this reason it is convenient to track the $B-L$
asymmetry and not the individual lepton or baryon asymmetry. 
The evolution of $B-L$ asymmetry is described by the following Boltzmann (rate) equation
\be\label{unflke}
{dN_{B-L}\over dz}  = 
\sum_{i=1}^3\,\varepsilon_i\,D_i\,(N_{N_i}-N_{N_i}^{\rm eq})-
N_{B-L}\,[\D W(z)+\sum_i \,W_i^{\rm ID}(z)] \;  ,
\ee
that can be solved as a coupled equation together 
with the Boltzmann equations for the RH neutrino abundances eqs.~(\ref{BERHNabundances}). 
The first term in the right-hand side is a source term for the asymmetry
and is proportional to
the three total $C\!P$  asymmetries  defined as 
\be
\ve_i \equiv - {\G_i-\bar{\G}_i \over \G_i+\bar{\G}_i} \,  ,
\ee
corresponding to the $B-L$ asymmetry produced, on average, by each single $N_i$ decay.
A perturbative calculation 
\setcounter{footnote}{17}\footnote{In quantum field theory a `perturbative calculation'
of the  scattering amplitudes and of the decay rates is possible when the interaction is sufficiently weak.
In this case one can use a perturbation expansion based on the Dyson expansion
for the  so called $S$ matrix \cite{mandlshaw}, the matrix that evolves the initial state
in some elementary process under consideration into the final  state. The different terms in the perturbation expansion 
can  be represented in a powerfully compact diagrammatic way in terms of the Feynamn diagrams.
In our case the interaction is given by the Yukawa interaction $h$ in the lagrangian eq.~(\ref{lagrangian}).
The validity of the perturbative calculation holds  when $h\lesssim 1$, as it is usually assumed in models.}
from the interference of tree level with one
loop self-energy and vertex diagrams (see Fig.~4),  gives \cite{crv}
\be\label{CPas}
\ve_i =\, {3\over 16\pi}\, \sum_{j\neq i}\,{{\rm
Im}\,\left[(h^{\dagger}\,h)^2_{ij}\right] \over
(h^{\dagger}\,h)_{ii}} \,{\xi(x_j/x_i)\over \sqrt{x_j/x_i}}\, ,
\ee
\begin{figure}
\begin{center}
\begin{minipage}{150mm}
\begin{center}\vspace*{-60mm}
\subfigure{\resizebox*{120mm}{!}{\includegraphics{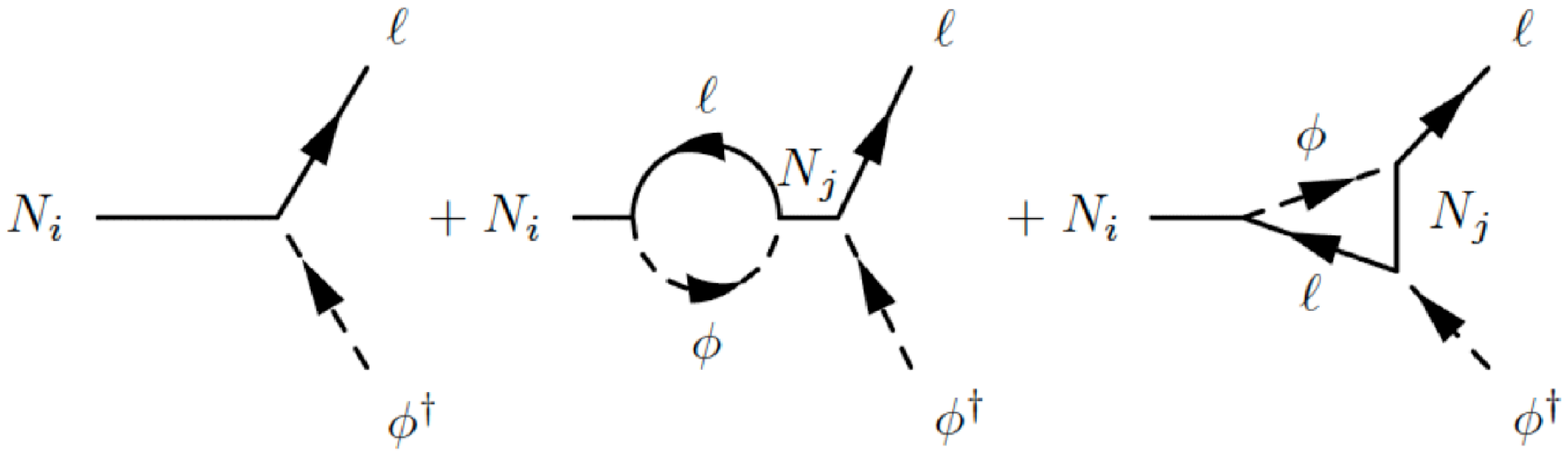}}}
\vspace*{-65mm}
\caption{The Feynman diagrams that are employed in the calculation  of 
the expression of the $C\!P$ asymmetry  eq.~(\ref{CPas}):  
tree level diagram (left), one loop self energy diagram (centre), one loop vertex diagram (right).}
\end{center}
\end{minipage}          
\end{center}
\end{figure}
where
\be\label{xi}
\xi(x)= {2\over 3}\,x\,
\left[(1+x)\,\ln\left({1+x\over x}\right)-{2-x\over 1-x}\right]  \, .
\ee
If it were not for the interference, in a pure tree level calculation,
one would obtain $\G_i = \bar{\G}_i$. 

The second term in the right-hand side of the eq.~(\ref{unflke}) is the so called wash-out term.
It is a term that tends to re-equilibrate the number of leptons and anti-leptons
destroying the asymmetry created by the source $C\!P$ violating term.
Differently from that term, it does not depend at all on the $C\!P$
asymmetries and it is present even if these vanish. It is simply
a statistical re-equilibrating term that has to be present in order
for the third Sakharov condition, on the departure from thermal equilibrium,
to be respected.  This is because if there is no (permanent) departure from thermal equilibrium,
this term would destroy completely any asymmetry previously generated.

The wash-out term is proportional to the wash-out factor that is
composed of two terms: a  term $\sum_i W_i^{\rm ID} (z)$ due to inverse decays 
(i.e. ${\ell}_i \, (\bar{\ell}_i)+\phi^{\dagger} \, (\phi)\ra N_i$)
and a  term  $\D W(z)$ due to (non-resonant) $\Delta L=2$ processes 
(${\ell}_i + \phi^{\dagger} \leftrightarrow \bar{\ell}_i + \phi $). 
After proper subtraction of the resonant contribution from
$\Delta L=2$ processes, the inverse decay
washout terms are simply given by
\be\label{WID}
W_i^{\rm ID}(z) =
{1\over 4}\,K_i\,\sqrt{x_i}\,{\cal K}_1(z_i)\,z_i^3 \,  ,
\ee
 where $z_i\equiv z\,\sqrt{x_i}$.
The washout term $\D W (z)$ gives a non-negligible effect only at $z\gg 1$ and in this case
it can be approximated as \cite{upperbound}
\be
\Delta W(z) \simeq {\o \over z^2}\,\left(M_1\over 10^{10}\,{\rm GeV}\right)\,
\left({\overline{m}^{\, 2} \over {\rm eV^2}}\right) \, ,
\ee
where $\o\simeq 0.186$ and $\overline{m}^2\equiv m_1^2+m_2^2+m_3^2$.

The solution of the Boltzmann equations for the final $B-L$ asymmetry can be written
as the sum of two contributions,
\be\label{NBmLf}
N^{\rm f}_{B-L} = N^{\rm pre-ex,f}_{B-L}+ N^{\rm lep,f}_{B-L}\, .
\ee
The first term, $N^{\rm pre-ex,f}_{B-L}$, is the residual value of a possible 
pre-existing asymmetry generated by some external mechanism prior to
the onset of leptogenesis.  Therefore, it represents a possible external contribution whose initial
value would originate from some source beyond leptogenesis. Therefore, if one wants
to explain the observed asymmetry with leptogenesis, this term has to be negligible.

The second term is the genuine leptogenesis contribution to the final asymmetry and
is the sum of three contributions, one for each RH neutrino species,
\be
N^{\rm lep,f}_{B-L}= \sum_i \, \ve_i \, \kappa_i^{\rm f}(K_1, K_2, K_3) \, .
\ee
Each contribution is the product of the total $C\!P$ asymmetry $\ve_i$  
times  the final value of the efficiency factor, $\k_i^{\rm f}(K_i)$, 
depending on  the decay parameter $K_i$.
The efficiency factors can be simply calculated analytically as
\be\label{efial}
\k_{i}(z;K_i)=-\int_{z_{\rm in}}^z \,dz'\,{dN_{N_i}\over dz'}\,
{\rm e}^{- z'\, \D W(z')+\sum_i\,\int_{z'}^\infty\,dz''\,W_i^{\rm ID}(z'';K_i)]} 
\ee
and their final values $\k_i^{\rm f}(K_i)=\k_{i}(z=\infty;K_i)$.
Finally, the predicted baryon-to-photon ratio $\eta_B$ can be calculated
from  the final value of the final $B-L$ asymmetry using the relation
\be\label{etaB}
\eta_B^{\rm lep}=a_{\rm sph} {N_{B-L}^{\rm lep,f}\over N_{\g}^{\rm rec}}\simeq 0.96\times
10^{-2} N_{B-L}^{\rm f}\, ,
\ee
where $N_{\g}^{\rm rec}\simeq 37$, and $a_{\rm sph}=28/79 \simeq 1/3$.
The factor $1/N_{\g}^{\rm rec}$ takes into account the photon production 
after leptogenesis, due to the annihilations of SM particles, 
that has the effect to dilute the asymmetry with respect to the number of photons. 
The factor $a_{\rm sph}$ is the fraction of the $B-L$ asymmetry ending up into a 
baryon asymmetry under the action of sphaleron processes. 
The condition of successful leptogenesis  corresponds to
impose that $\eta_B^{\rm lep}$ reproduces the measured value of $\eta_B^{CMB}$  (cf. eq.~(\ref{etaBCMB})). 

 There is a  second important assumption within the vanilla scenario in the calculation
 of the baryon asymmetry: the RH neutrino mass spectrum is assumed to be hierarchical
with  $M_2 \gtrsim 3\, M_1$. 
Together with the  assumption of unflavoured leptogenesis,
the final asymmetry  is in this case  
dominantly produced by the lightest RH neutrino out-of-equilibrium decays,
in  a way that the sum in the eq.~(\ref{NBmLf}) can be approximated by
the first term ($i=1$), explicitly
\be
N_{B-L}^{\rm f}\simeq \ve_1\,\k_1^ {\rm f}(K_1)    \,  ,
\ee
and a `$N_1$-dominated scenario' holds.  In Fig.~3 we show 
a calculation of $\k^{\rm f}(K_1)$ for $K_1=0.01$ (weak wash-out regime)
and for $K_1 = 10$ (strong wash-out regime) and for initial thermal abundance
($N_{N_1}^{\rm in}=1$).  In the first case all the
asymmetry produced in the decays of the RH neutrino survives and $\k_1^{\rm f}=N^{\rm in}_{N_1}=1$.
In the second case only a small fraction of the asymmetry survives and $\k_1^{\rm f}\simeq 0.01 $,
still however large enough to reproduce successfully the observed asymmetry if $\ve_1 \simeq 10^{-6}$.

The  `$N_1$-dominated scenario' holds  either because $|\ve_{2,3}|\ll |\ve_1|$ and/or because 
the asymmetry initially produced
by the $N_{2,3}$ decays is afterwards washed-out
by the lightest RH neutrino inverse processes in a way that
$\k_{2,3}^{\rm f} \ll \k_1^{\rm f}$.  Indeed
if we indicate with $N_{B-L}^{(2,3)}(T\gtrsim M_1)$ the contribution
to the $N_{B-L}$ asymmetry from the two heavier RH neutrinos prior
to the lightest RH neutrino wash-out, the final values are given very simply by
\be \label{N1washunfl}
N_{B-L}^{(2,3), {\rm f}} = N_{B-L}^{(2,3)}(T\gtrsim M_1) \, e^{-{3\pi \over 8}\,K_1} \, .
\ee
The same exponential wash-out factor applies to the residual value
of a possible pre-existing asymmetry. In this way it is sufficient 
to have a strong wash-out condition $K_1 \gg 1$ 
in order to have both  a pre-existing asymmetry 
and a contribution from heavier RH neutrinos  negligible. The strong wash-out
condition $K_1 \gg 1$ is very easily satisfied since, barring special cases, one has
typically $K_1 \simeq (m_{\rm sol}\div m_{\rm atm})/10^{-3}\,{\rm eV} \gg 1$.
The same condition  also guarantees independence of the final
asymmetry on the initial $N_1$-abundance. It is then quite suggestive that
the measured values of $m_{\rm sol}$ and $m_{\rm atm}$ have just the right values
to produce a wash-out that is strong enough to guarantee independence on the initial
conditions but still not too strong to prevent successful leptogenesis. This leptogenesis
conspiracy between experimental results and theoretical prediction is one of the main reasons
that has determined the success of leptogenesis so far. 

There is actually a particular case where $K_1 \gg 1$ and $|\ve_{2}|\ll |\ve_1|$ do not hold and in this case
the final asymmetry is dominated by the contribution coming from 
the next-to-lightest  RH neutrinos.  
However, one still has $K_2 \gg 1$ so that
the independence of the initial conditions still holds. 
For the time being, as an additional
third assumption of the vanilla scenario, we will bar this particular case,
we will be back on it in 3.1.

If, additionally, one excludes fine tuned cancelations among
the different terms contributing to the neutrino masses 
in the see-saw formula, one obtains the following upper bound 
on the lightest RH neutrino $C\!P$ asymmetry \cite{di}
\be\label{CPbound}
\ve_1 \leq \ve_1^{\rm max} 
\simeq 10^{-6}\,{M_1\over 10^{10}\,{\rm GeV}}\,{m_{\rm atm}\over m_1+m_3} \, .
\ee
Imposing $\eta_B^{\rm max}\simeq 0.01\,\ve_1^{\rm max}\,\k_1^{\rm f} > \eta_{B}^{CMB}$,
one obtains the allowed region in the plane $(m_1,M_1)$ shown in the left panel of Fig.~5.
\begin{figure}
\begin{center}
\begin{minipage}{100mm}
\subfigure[]{
\resizebox*{5cm}{!}{\includegraphics{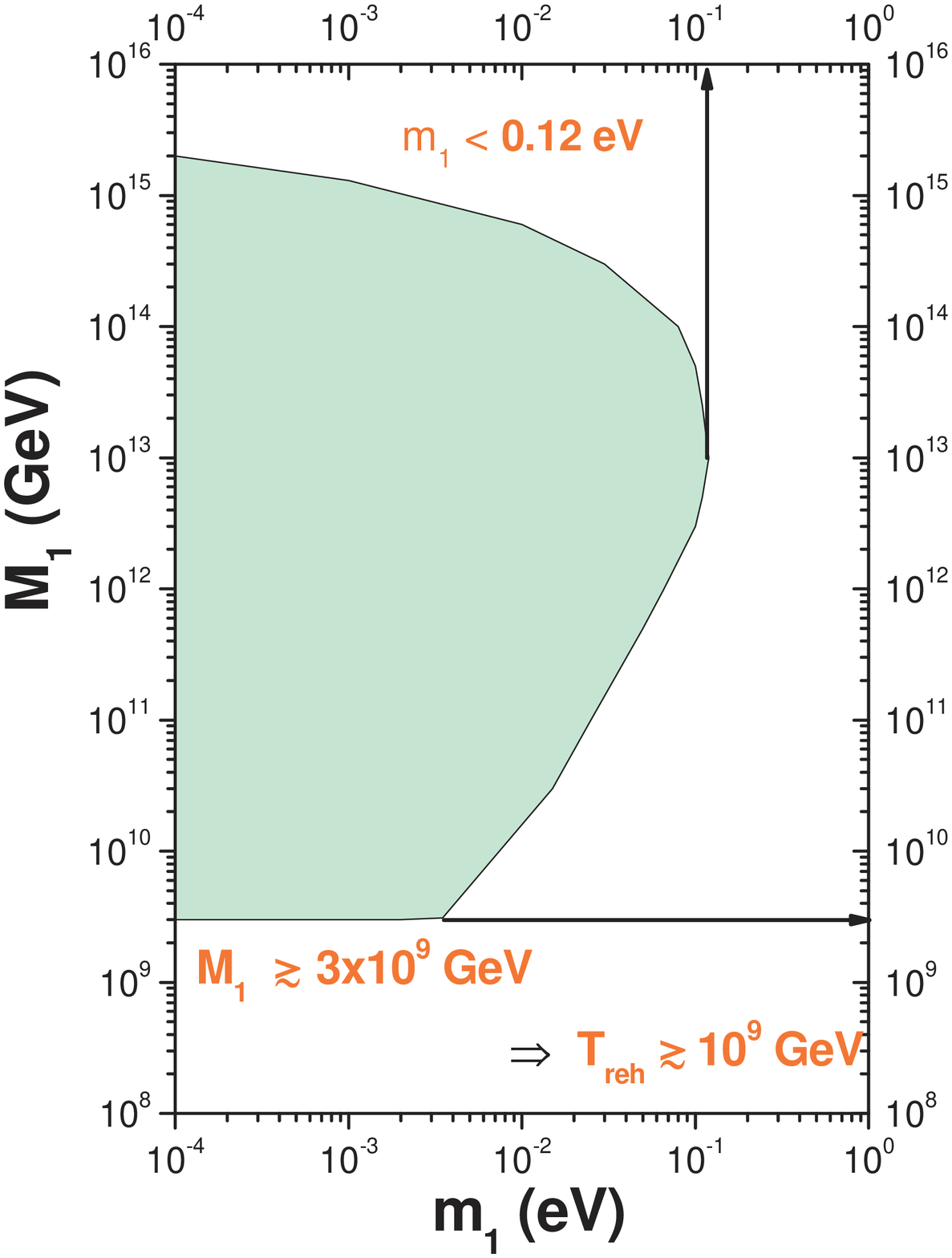}}}
\subfigure[]{\vspace*{-20mm}
\resizebox*{45mm}{!}{\includegraphics{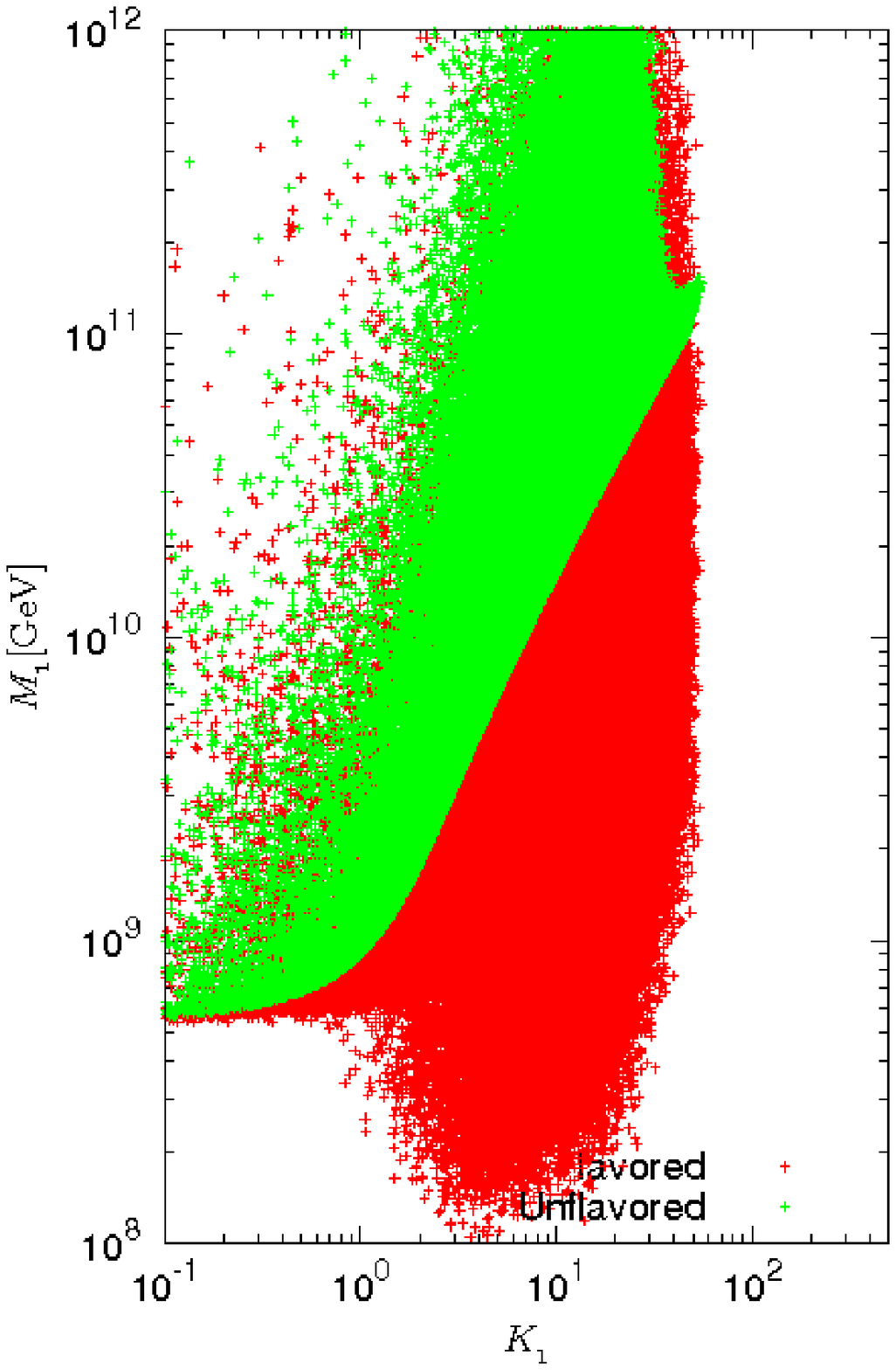}}}
\caption{Left: Neutrino mass bounds in the vanilla scenario. Right:
         Relaxation of the lower bound on $M_1$ thanks
          to additional unbounded flavoured $C\!P$ violating terms.}
\end{minipage}          
\end{center}
\end{figure}
One can notice the existence of an upper bound on the light
neutrino masses $m_1\lesssim 0.12\,{\rm eV}$ \cite{upperbound}, 
incompatible with quasi-degenerate
neutrino mass models, and a lower bound on
$M_1\gtrsim 3\times 10^9\,{\rm GeV}$ \cite{di,upperbound} 
implying a lower bound on the
reheat temperature $T_{\rm RH}\gtrsim 10^9\,{\rm GeV}$ \cite{upperbound}.

An important feature of vanilla leptogenesis is that the final asymmetry does not
directly depend on the parameters of leptonic mixing matrix $U$. 
This implies that one cannot establish a direct model independent connection. 
In particular a discovery
of CP violation in neutrino mixing would not be a smoking gun for leptogenesis
and, vice versa, a non-discovery would not rule out leptogenesis.
However, within more restricted scenarios, for example imposing some conditions on
the neutrino Dirac mass matrix, links can emerge. We will discuss
in detail the interesting case of $SO(10)$-inspired models.

\section{Beyond vanilla leptogenesis}

In  the last years different directions beyond the
vanilla leptogenesis scenario have been explored, 
usually aiming at evading the above-mentioned bounds. 
The most important one is certainly the inclusion of
flavour effects and for this reason we will discuss 
them in detail in the next Section.  
In this section  we briefly mention other important 
extensions of the vanilla scenario 
that have been considered and studied. 

\subsection{Beyond a hierarchical RH neutrino mass spectrum}

If $(M_2-M_1)/M_1 \equiv \delta_2 \ll 1$, the $C\!P$ asymmetries $\ve_{1,2}$ get resonantly enhanced as $\ve_{1,2}\propto 1/\d_2$ \cite{crv}. 
If, more stringently, $\d_2\lesssim 10^{-2}$, then
$\eta_B \propto 1/\delta_2$ and the degenerate limit is obtained \cite{beyondHR}.
In this limit the lower bounds on $M_1$ and on $T_{\rm RH}$
get relaxed $\propto \delta_2$ and at the resonance they completely disappear 
\cite{resonant}.
However, there are not many models able to justify in a reasonable way such a very 
closely degenerate limit yielding resonant leptogenesis.

\vspace*{-3mm}
\subsection{Non minimal leptogenesis}

Other proposals to relax the lower bounds on $M_1$ and on $T_{RH}$
rely on extensions  beyond minimal leptogenesis. For example a popular
scenario relies on the
addition of a so called left-left Majorana mass term to the lagrangian (\ref{lagrangian}).
However, this implies that the existence of new particles, beyond the
SM particles and  RH neutrinos, has to be postulated
since otherwise it would not be possible to write such a term
preserving the symmetries of the lagrangian. The most popular way is to include a
new kind of Higgs boson, $\Delta$ with mass $M_{\D}$, 
with different properties than the SM Higgs boson. In this way one can show that in the
see-saw formula eq.~(\ref{see-saw}) for the neutrino masses 
a second term appears. This term is $\propto (v^2/M_{\D})\times I$, 
where $I$ is the identity matrix. This extension of the seesaw mechanism is called   
type II see-saw mechanism \cite{typeII} and it is particularly suitable
to describe quasi-degenerate neutrino masses. However, as shown in Fig.~1, 
these start to be strongly constrained by cosmology and a simple 
type II extension does not seem strictly necessary within the current 
experimental results. 

Another popular way to go beyond a minimal scenario of leptogenesis 
relies on a non thermal production of the RH neutrinos generating 
the asymmetry \cite{nonth}.
However, these non minimal models spoil somehow the remarkable coincidence 
between the measured values of the atmospheric and solar neutrino mass scales 
and the possibility to have successful leptogenesis with $\eta_B\sim 10^{-9}$
even independently of the initial conditions.
Non-minimal models have been also extensively recently explored in order to get a
low scale leptogenesis testable at colliders \cite{see-sawLHC}.

\vspace*{-2mm}
\subsection{Improved kinetic description}

Within vanilla leptogenesis the asymmetry is calculated solving simple rate equations,
classical Boltzmann equations integrated over the RH neutrino momenta.
Different kinds of extensions have been studied \cite{momentum}, for example accounting 
for $\Delta L=1$ scatterings,  momentum dependence, quantum kinetic effects 
and thermal effects.
All these analyses find  significant changes in the weak wash-out regime but within
${\cal O}(1)$ in the strong wash-out regime. This result has
quite a straightforward general explanation \cite{upperbound}. In the strong wash-out regime the final asymmetry
is produced by the decays of RH neutrinos in a non relativistic regime 
when a simple classical momentum independent kinetic description provides quite a good approximation.
The use of a simple kinetic description in leptogenesis
is therefore not just a simplistic approach but is justified in terms of the
neutrino oscillation experimental results on the neutrino masses that
support a strong wash-out regime with $K_i \gg 1$.

\section{The importance of flavour}

In the last years, the inclusion of flavour effects in the calculation
of the final asymmetry  proved to yield the most significant
changes to the calculation of the final asymmetry
going beyond the vanilla scenario. 

There are two kinds of flavour effects that are neglected in the vanilla scenario: heavy
neutrino flavour effects \cite{geometry}, how heavier RH neutrinos contribute
to the final asymmetry, and lepton flavour effects \cite{flavoreffects1,flavoreffects2}, 
how the
flavour composition of the leptons quantum states produced in the RH neutrino decays
affects the calculation of the final asymmetry.
We first discuss the two effects separately and then we show
how their interplay has very interesting consequences.

\subsection{Heavy neutrino flavour effects}

In the vanilla scenario the contribution to the final asymmetry from the
heavier RH neutrinos is negligible because either the $C\!P$ asymmetries  are
suppressed in the hierarchical limit with respect to $\ve_1^{\rm max}$ (cf. eq.~(\ref{CPbound})) and/or because
even assuming that a sizeable asymmetry is produced around $T \sim M_{2,3}$,
this is later on washed out by the lightest RH neutrino inverse processes.

However, as we anticipated, there is a particular
case  when, even neglecting the lepton flavour composition and assuming
a hierarchical heavy neutrino mass spectrum,  the contribution
to the final asymmetry from next-to-lightest RH neutrino decays can be 
dominant \cite{geometry}.  This case corresponds to a particular 
choice of the orthogonal matrix such that $N_1$ is very weakly coupled 
(corresponding to have $K_1 \ll 1$) and its wash-out can be neglected.
For the same choice of the parameters, the $N_2$ total $C\!P$ asymmetry 
$\ve_2$ is unsuppressed if $M_3\lesssim 10^{15}\,{\rm GeV}$.
In this case a $N_2$-dominated scenario is realised.
Notice that in this case the existence of a third heaviest RH neutrino 
species is crucial in order to have a sizeable $\ve_2$. 

The contribution from the next-to-lightest RH neutrino species  $N_2$
is also important  in the quasi-degenerate limit when $\d_{2} \ll 1$. In this
case the $C\!P$ asymmetries $\ve_{2}$
is also not suppressed and the wash-out
from the lightest RH neutrinos on the asymmetry produced by $N_2$
is only partial \cite{resonant,beyondHR}. 
Analogously, if $\d_3 \equiv (M_3 - M_1)/M_1 \ll 1$, then
the contribution to the final asymmetry from $N_3$ has also in general
to be taken into account. 

\subsection{Lepton flavour effects}

For the time being, let us continue to assume that the final asymmetry is
dominantly produced by the contribution of the lightest RH neutrinos $N_1$,
neglecting the contribution from the decays of the heavier RH neutrinos $N_2$ and $N_3$.
If $M_1\gg 10^{12}\,{\rm GeV}$, the flavour composition
of the quantum states of the leptons produced in $N_1$ decays
has no influence on the final asymmetry and the `unflavoured regime', assumed
in the vanilla scenario, holds. 

This is because for such a large $M_1$, the asymmetry is produced at
temperatures where all the relevant interactions are flavour blind. 
In this way the lepton quantum states evolve coherently  between the production 
from a $N_1$-decay and a subsequent inverse decay with an Higgs boson and
the lepton flavour composition does not play any role in the calculation of the final asymmetry.

However, if $10^{12}\,{\rm GeV}\gtrsim M_1 \gtrsim 10^{9}\,{\rm GeV}$, 
during the relevant period of generation of the asymmetry, it can be shown that the rate 
of the tauon lepton interactions becomes much larger than the inverse decay rate.
As a consequence, the produced lepton quantum states on average, between one decay and a subsequent inverse decay, 
interact with tauons. In this way the tauon component of the lepton quantum
states is measured by the thermal bath and the coherent evolution breaks down. 
Therefore, at the subsequent inverse decay, 
the leptonic quantum states are an incoherent mixture of a tauon component and
of a (still coherent) superposition of an electron and of a muon component that 
we can indicate with ${\ell}_{e+\mu}$.
The fraction of the asymmetry stored in each flavour component is not proportional in general
to the branching ratio of that component. This implies that the two
flavour asymmetries, the tauon and the $e+\mu$ components, 
evolve differently and have to be calculated separately. 
In this way  the resulting final asymmetry can considerably differ
from the result one would obtain in the unflavoured regime. 
This regime  is called the `two fully flavoured regime'.

If $M_1\lesssim 10^{9}\,{\rm GeV}$, then even the coherence of the $e+\mu$
component is broken by the muon interactions between decays and inverse decays during the
relevant period of generation of the asymmetry. 
In this situation  all three flavoured asymmetries (electron, muon and tauon asymmetries)
evolve independently of each other 
and a `three fully flavoured regime' applies. In the intermediate regimes
a density matrix formalism is necessary to properly describe  decoherence  
\cite{flavoreffects1,densitymatrix,densitymatrix2}.

It is useful to discuss explicitly how the calculation of the baryon asymmetry
proceeds in the three fully flavoured regime. Defining the flavoured
asymmetries as $\D_{\a} \equiv B/3 - L_{\a}$, their abundances $N_{\D_{\a}}$, 
obeying $\sum_{\a=e,\m,\t} N_{\D_{\a}}= N_{B-L}$,  can be calculated solving the following  
Boltzmann equations 
\be\label{flke}
{dN_{\D_\a}\over dz}  = \ve_{1\a}\,D_1\,(N_{N_1}-N_{N_1}^{\rm eq})-
p_{1\a}^0\,N_{\D_{\a}}\,W_1^{\rm ID}(z) \hspace{10mm}  (\a=e,\m,\t) \;  ,
\ee
that has to be solved together 
with the Boltzmann equation for $N_{N_1}$ (cf. eq.~(\ref{BERHNabundances})). 
In the first term of the right-hand side, we have a source term for the flavoured asymmetry 
that is proportional to the the flavoured $C\!P$  asymmetry $\ve_{1\a}$.
In general, for a given RH neutrino $N_i$, the flavoured $C\!P$ asymmetries are defined as 
\be
\ve_{i\a} \equiv - {\G_{i\a}-\bar{\G}_{i\a} \over \G_i+\bar{\G}_i} \,  ,
\ee
where $\G_{i\a} \equiv p_{i\a}\,\G_i$ and $\bar{\G}_{i\a}\equiv \bar{p}_{i\a}\,\bar{\G}_i$
and $p_{i\a}$ and $\bar{p}_{i\a}$ are respectively  the branching ratios for the
decay of the RH neutrino $N_i$ into a lepton ${\ell}_{\a}$ and into an anti-lepton $\bar{\ell}_{\a}$. 

The wash-out factor is smaller than in the unflavoured case since $W_1^{\rm ID}(z)$
is now replaced by $p^0_{1\a}\,W_1^{\rm ID}(z)$, where $p^0_{1\a}$ is 
the average of $p_{1\a}$ and $\bar{p}_{1\a}$.  Correspondingly, it will also prove useful to define the flavoured
decay parameters $K_{i\a}\equiv p^0_{i\a}\,K_i$. 

The final $B-L$ asymmetry has now to be calculated as the sum of the final values of the flavoured asymmetries,
$N^{\rm f}_{B-L}=\sum_{\a}\,N^{\rm f}_{\D_{\a}}$.
Even though the kinetic equations for the $N_{\D_{\a}}$ (cf. eq.~(\ref{flke})) are 
analogous to the eq.~(\ref{unflke}) for the calculation of $N_{B-L}$
in the unflavoured regime,  the final result for $\eta_B$ can be in general very different.   The most
drastic departure from the unflavoured results is the special case $\ve_1 =0$. In the vanilla scenario,
assuming the unflavoured regime, 
this would  simply yield $N^{\rm f}_{B-L}=\eta_B= 0$, i.e. no asymmetry would be produced at all. 
This is in agreement with an extension of the Sakharov condition to vanilla leptogenesis that implies
that $B-L$ has not  to be conserved. However, in the three fully flavoured regime, this condition
is too restrictive and it is enough that the $\D_{\a}$'s are not conserved in order to produced a non-vanishing
baryon asymmetry. 

The flavoured Boltzmann equations eqs.~(\ref{flke}), written in the three fully flavoured
regime, can be straightforwardly extended to the two fully flavoured regime: simply the equations
for $\D_e$ and $\D_{\m}$ are replaced by one equation for the asymmetry $\D_{e+\m}\equiv \D_e + \D_{\m}$,
where $\ve_{1e}$ and $\ve_{1\mu}$ are replaced by $\ve_{1,e+\m}\equiv \ve_{1e}+\ve_{1\m}$ 
and $K_{1e}$ and $K_{1\m}$ are replaced
by $K_{1,e+\m}\equiv K_{1e}+K_{1\m}$.
 
We can summarise the results for the calculation of the baryon asymmetry within the 
fully flavoured regimes (two or three) saying that
lepton flavour effects induce three major consequences.
i) The wash-out can be  considerably lower than in the unflavoured regime \cite{flavoreffects1,flavoreffects2}.
ii) The low energy phases affect directly the final asymmetry since they
contribute to a second source of $C\!P$ violation in the flavoured $C\!P$ asymmetries
\cite{flavorlep}. As a consequence the same source of $C\!P$
violation that could take place in neutrino oscillations, could be  sufficient
to explain the observed  asymmetry,
though under quite stringent conditions on the RH neutrino mass spectrum.
In the light of the recent overwhelming evidence of a non-vanishing $\theta_{13}$ angle 
\cite{theta13}, a necessary condition
to have $C\!P$ violation in neutrino oscillations, this problem becomes particularly interesting, in particular
a precise calculation, within a density matrix formalism, of the lower bound on $\theta_{13}$.
iii) Compared to the total $C\!P$ asymmetries, the flavoured $C\!P$ asymmetries contain
extra-terms that are not upper bounded if one allows for a  cancelations in
the see-saw formula among the light neutrino mass terms together with a mild
RH neutrino mass hierarchy ($M_2/M_1 \sim 10$). 
In this way the lower bound on $T_{RH}$
can be relaxed by about one order of magnitude, down to $10^8\,{\rm GeV}$
\cite{bounds} as shown in the right panel of Fig.~5.
However for most models, such as sequential dominated models \cite{sequential}, these cancellations do not occur.
In this case  lepton flavour effects cannot relax the lower bound on $T_{\rm RH}$.

\subsection{The interplay between lepton and heavy neutrino flavour effects}

As we have seen, when lepton flavour effects are neglected, the possibility
that the next-to-lightest RH neutrino decays contribute to the final
asymmetry is a special case. On the other hand,
when lepton flavour effects are taken into account, 
the contribution from next-to-lightest RH neutrinos 
cannot be neglected in general.  Even the contribution
from the heaviest RH neutrinos can be sizeable in some cases 
and has to be taken into account in general.
  
Assuming hierarchical mass patterns and that the RH neutrino processes 
occur in one of  the three fully flavoured regimes, there are
10 possible different mass patterns to consider (see Fig.~6) requiring specific multi-stage sets 
of classical Boltzmann equations for the calculation of the final asymmetry. 
\begin{figure}
\begin{minipage}{140mm}
\begin{center}
\subfigure[]{\resizebox*{27mm}{!}{\includegraphics{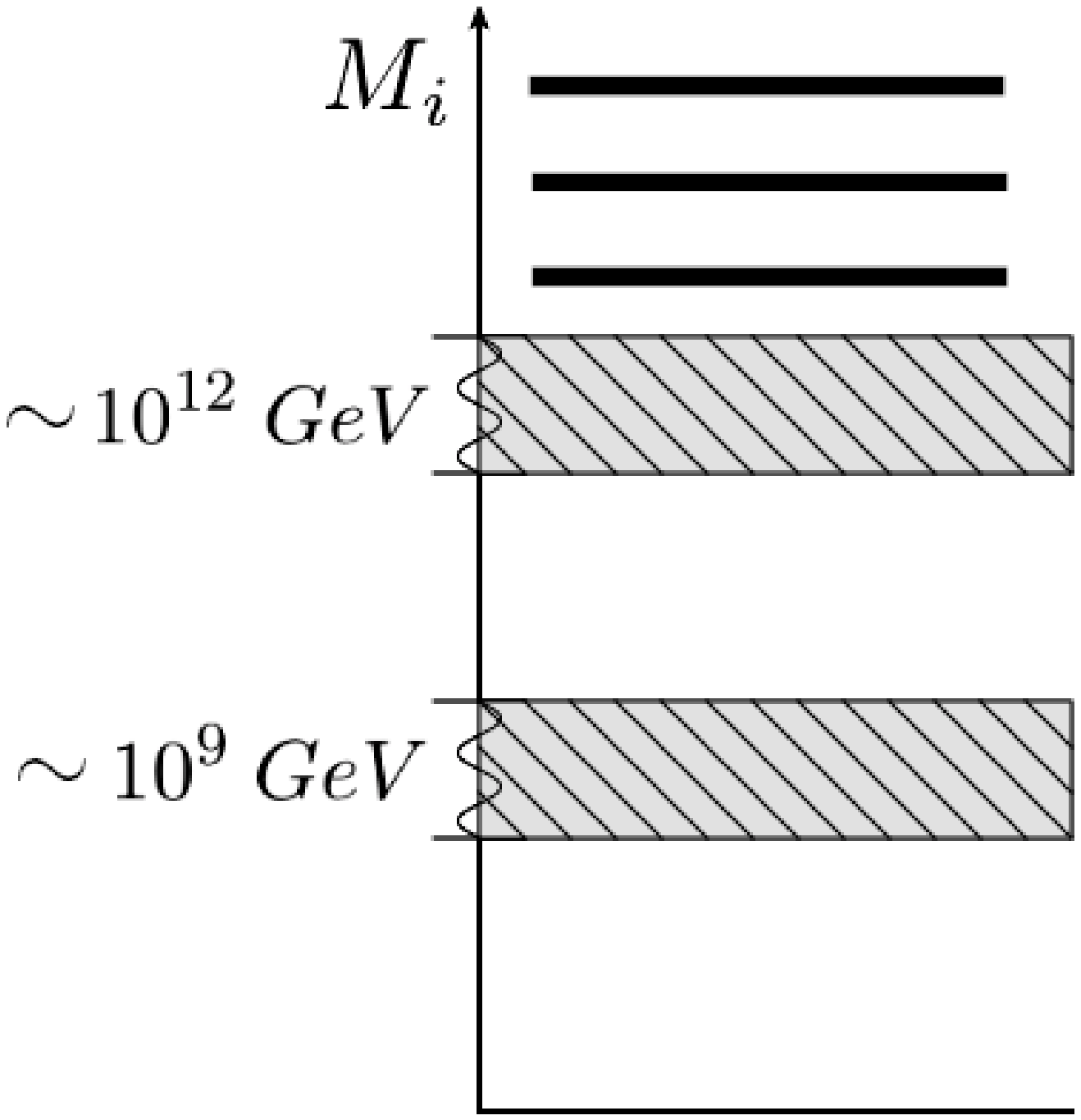}}} \hspace*{2mm}
\subfigure[]{\resizebox*{27mm}{!}{\includegraphics{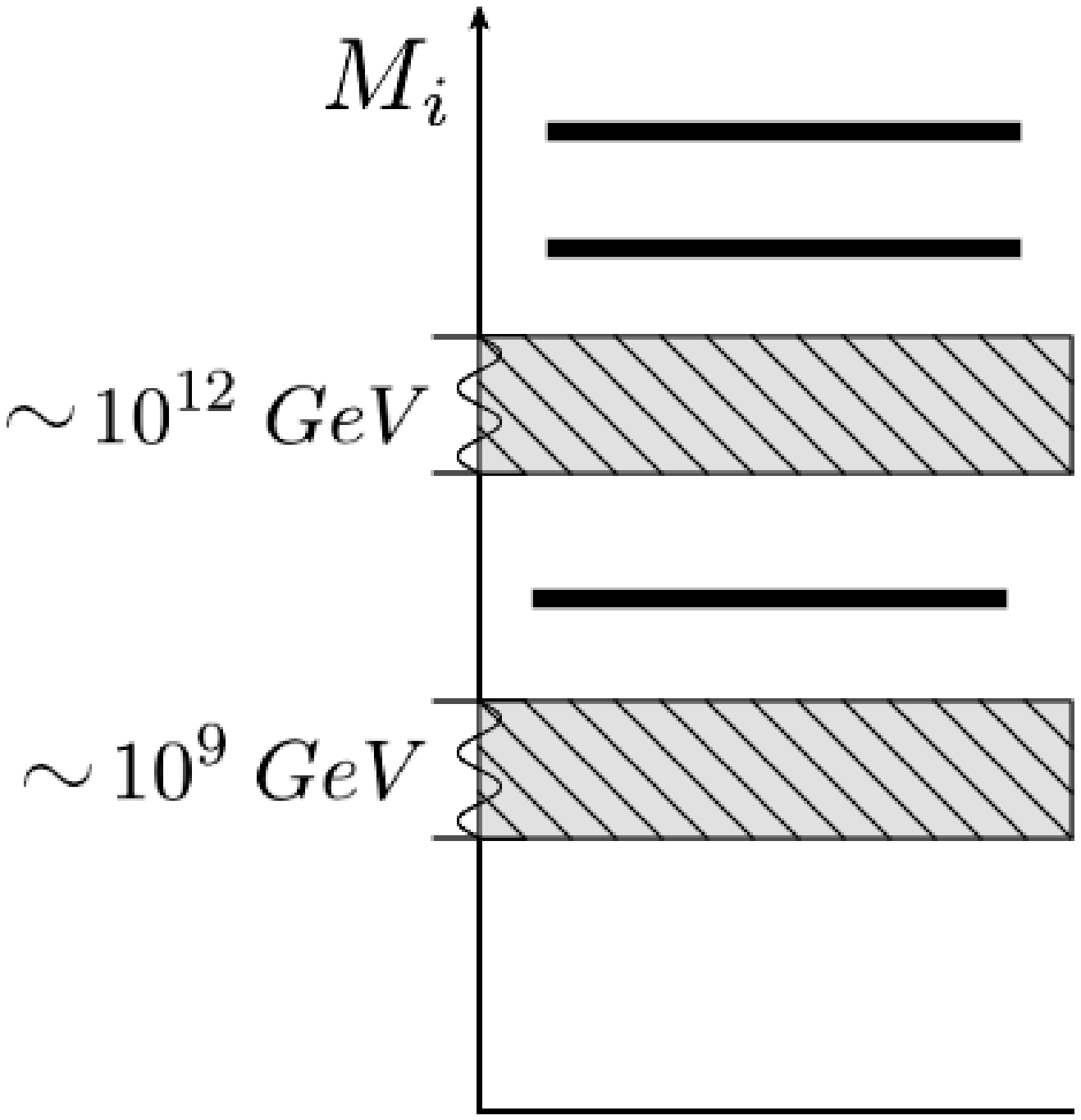}}} \hspace*{2mm}
\subfigure[]{\resizebox*{17mm}{!}{\includegraphics{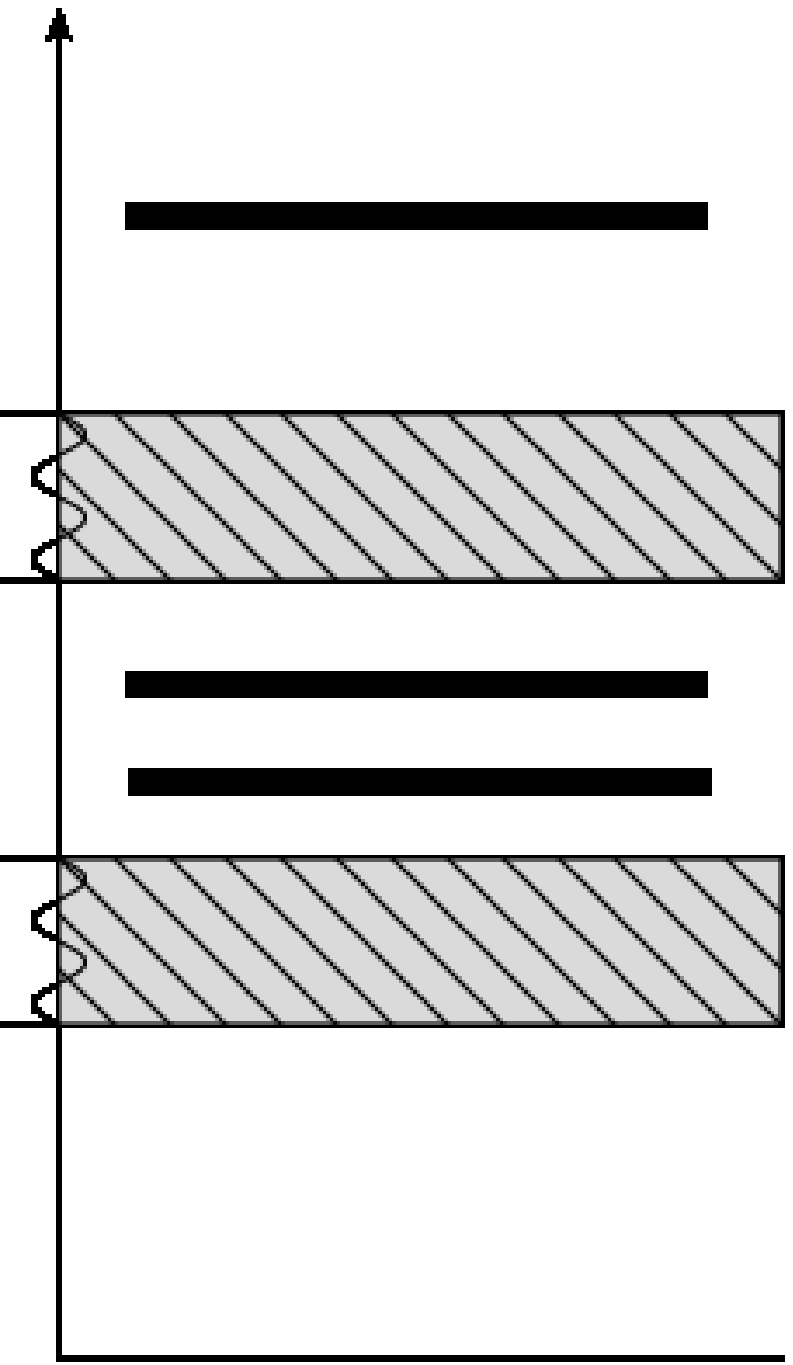}}} \hspace*{2mm}
\subfigure[]{\resizebox*{17mm}{!}{\includegraphics{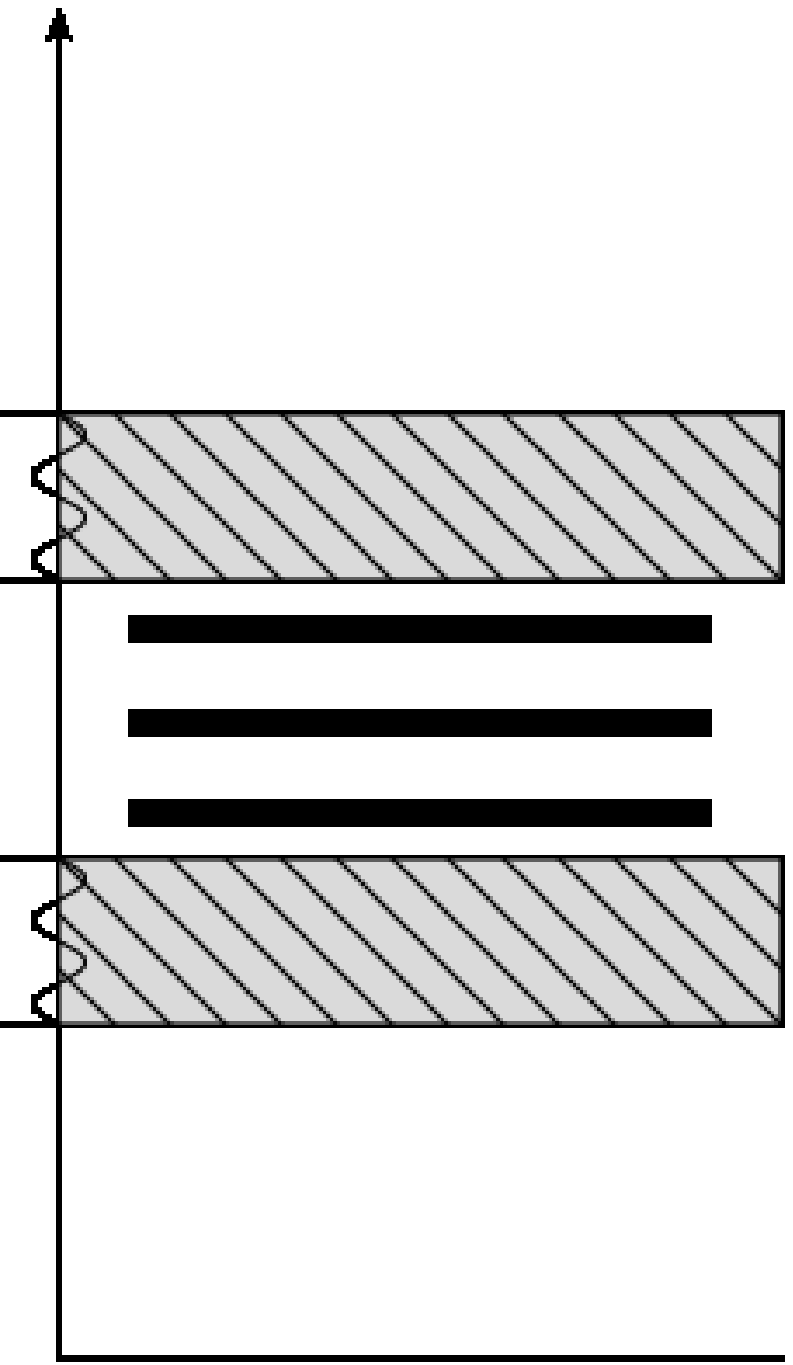}}} 
\end{center}
\begin{center}
\subfigure[]{\resizebox*{27mm}{!}{\includegraphics{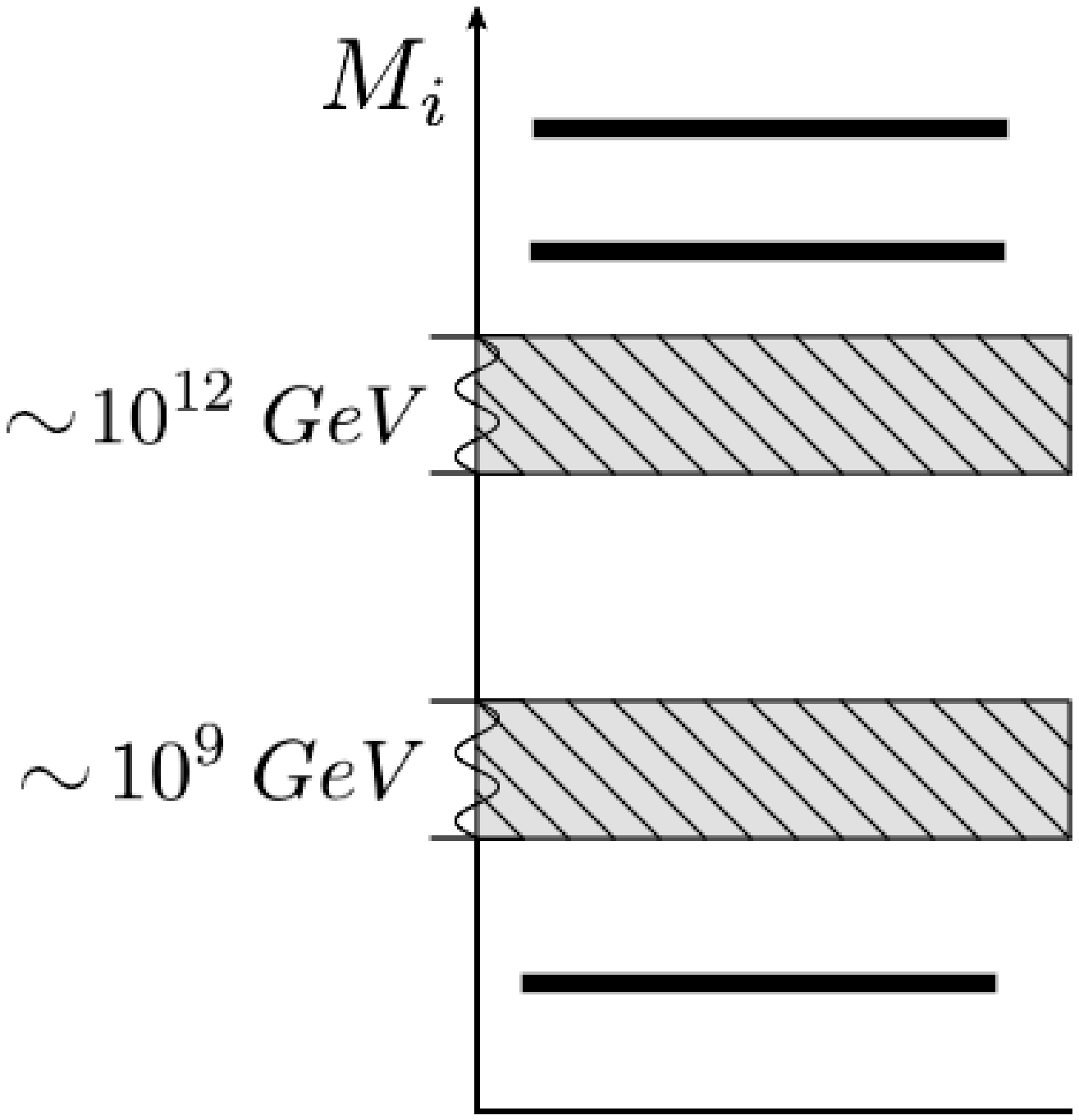}}}
\subfigure[]{\resizebox*{15mm}{!}{\includegraphics{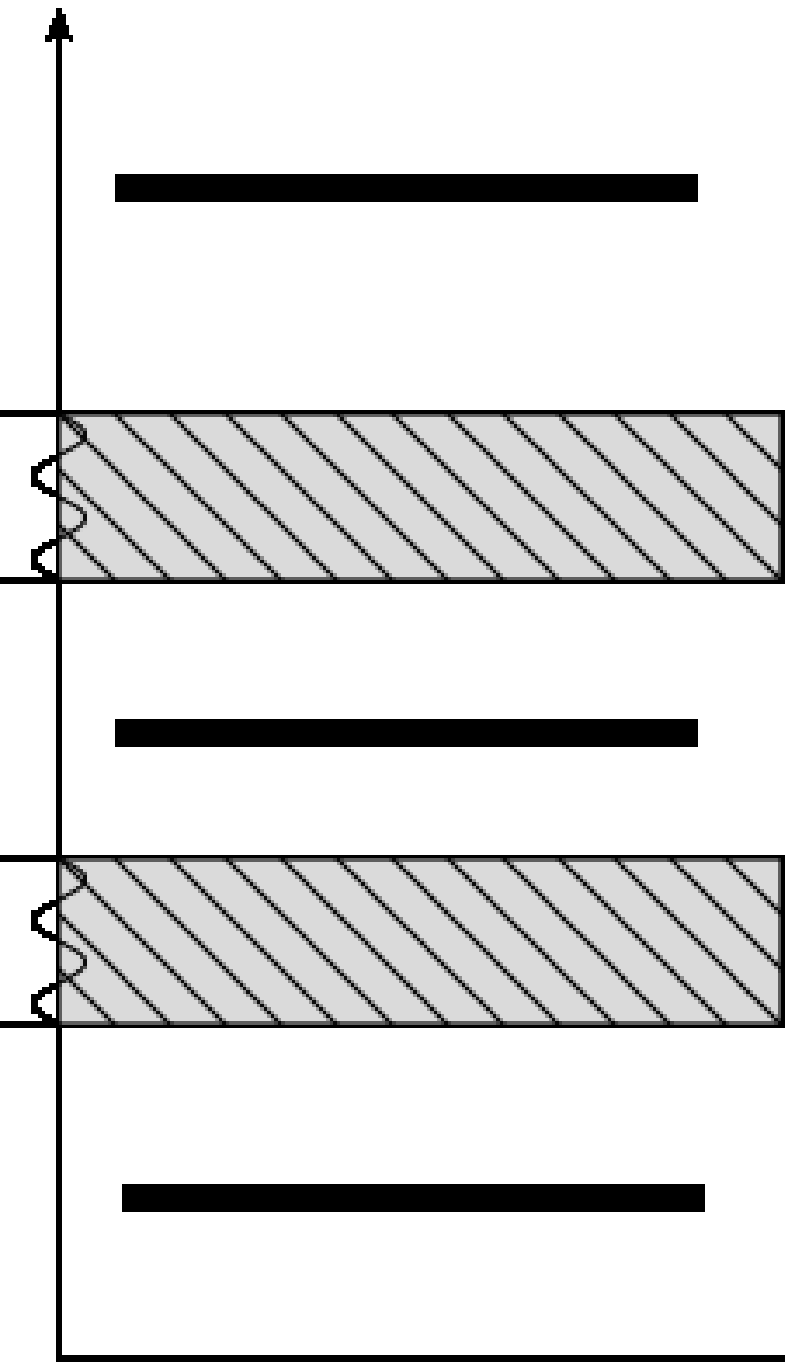}}}
\subfigure[]{\resizebox*{15mm}{!}{\includegraphics{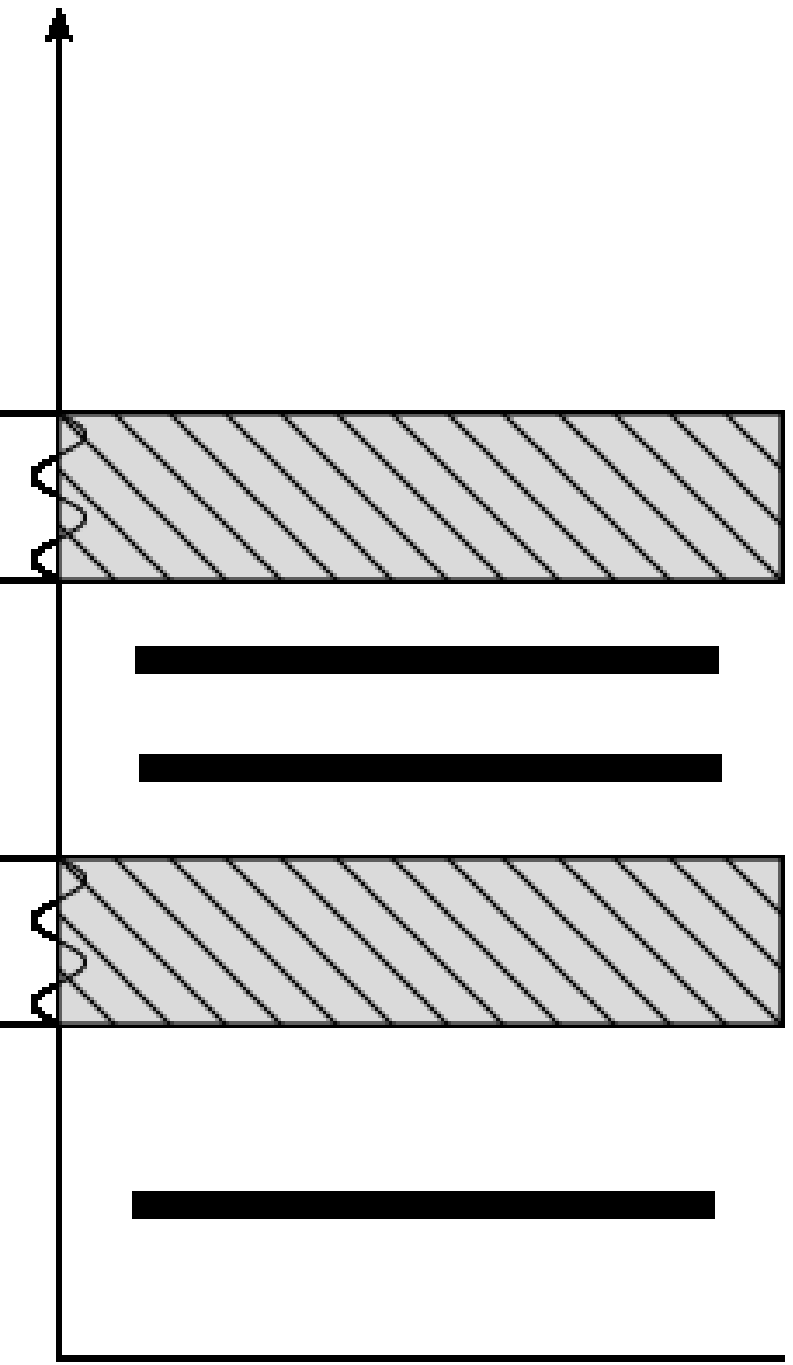}}} 
\subfigure[]{\resizebox*{30mm}{!}{\includegraphics{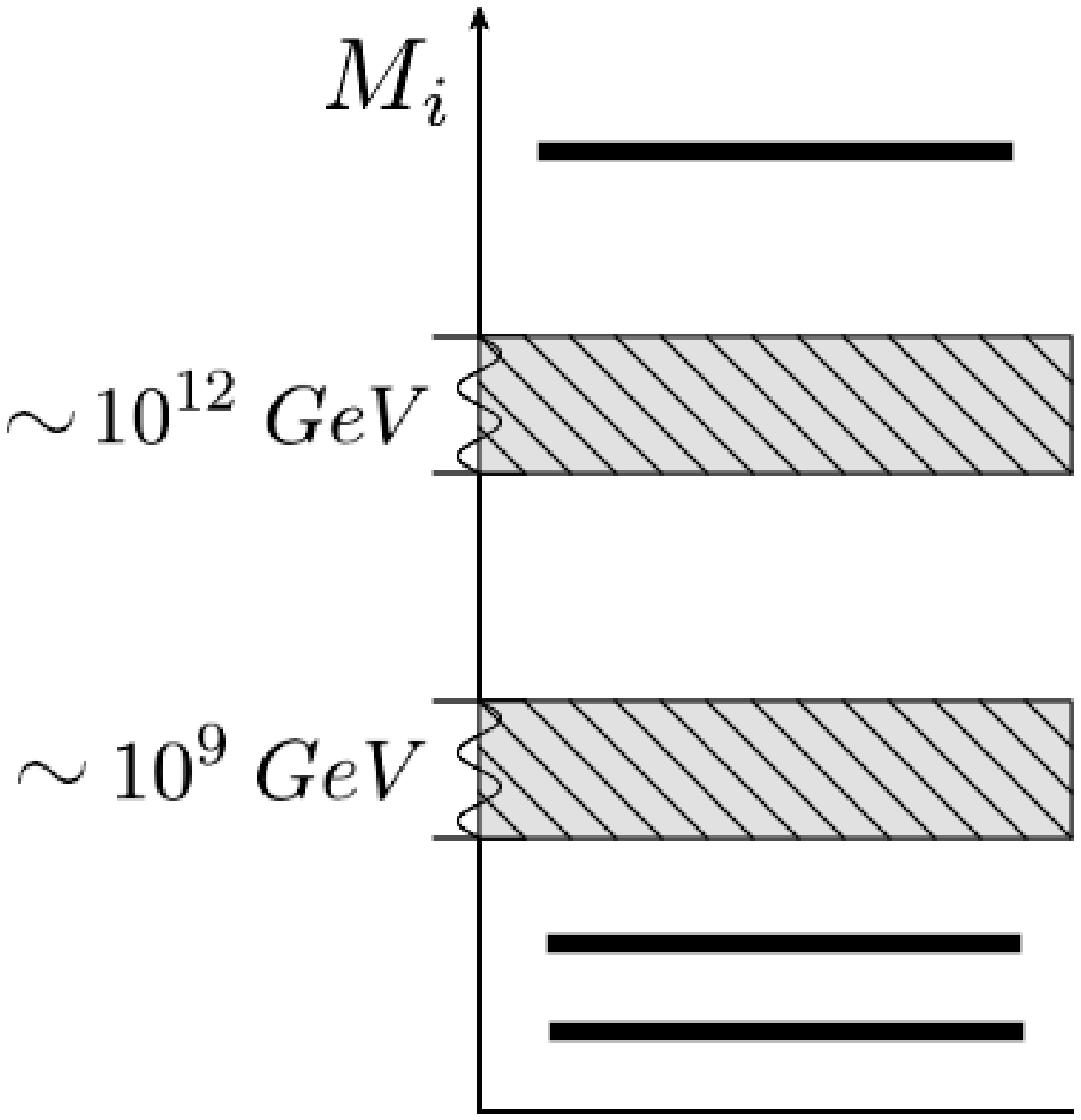}}}
\subfigure[]{\resizebox*{15mm}{!}{\includegraphics{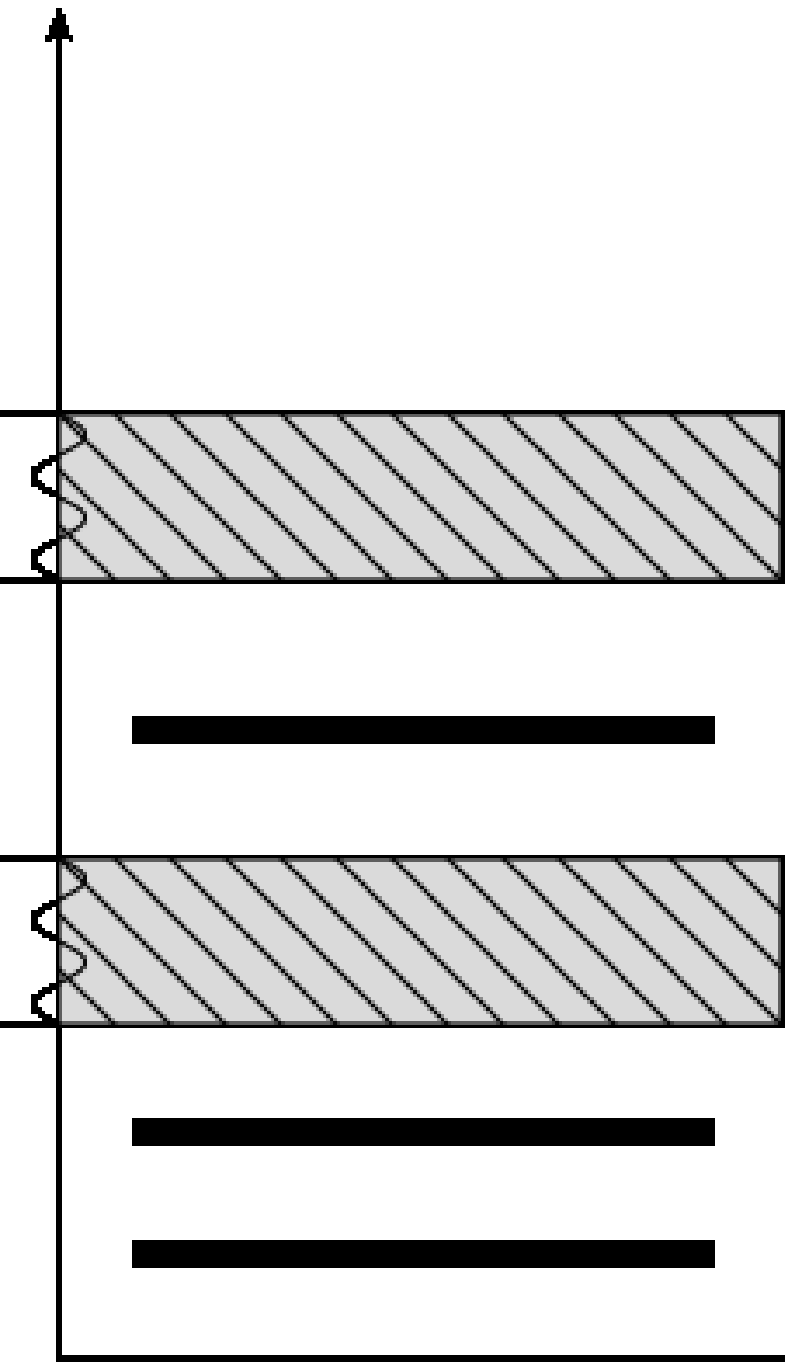}}}
\subfigure[]{\resizebox*{15mm}{!}{\includegraphics{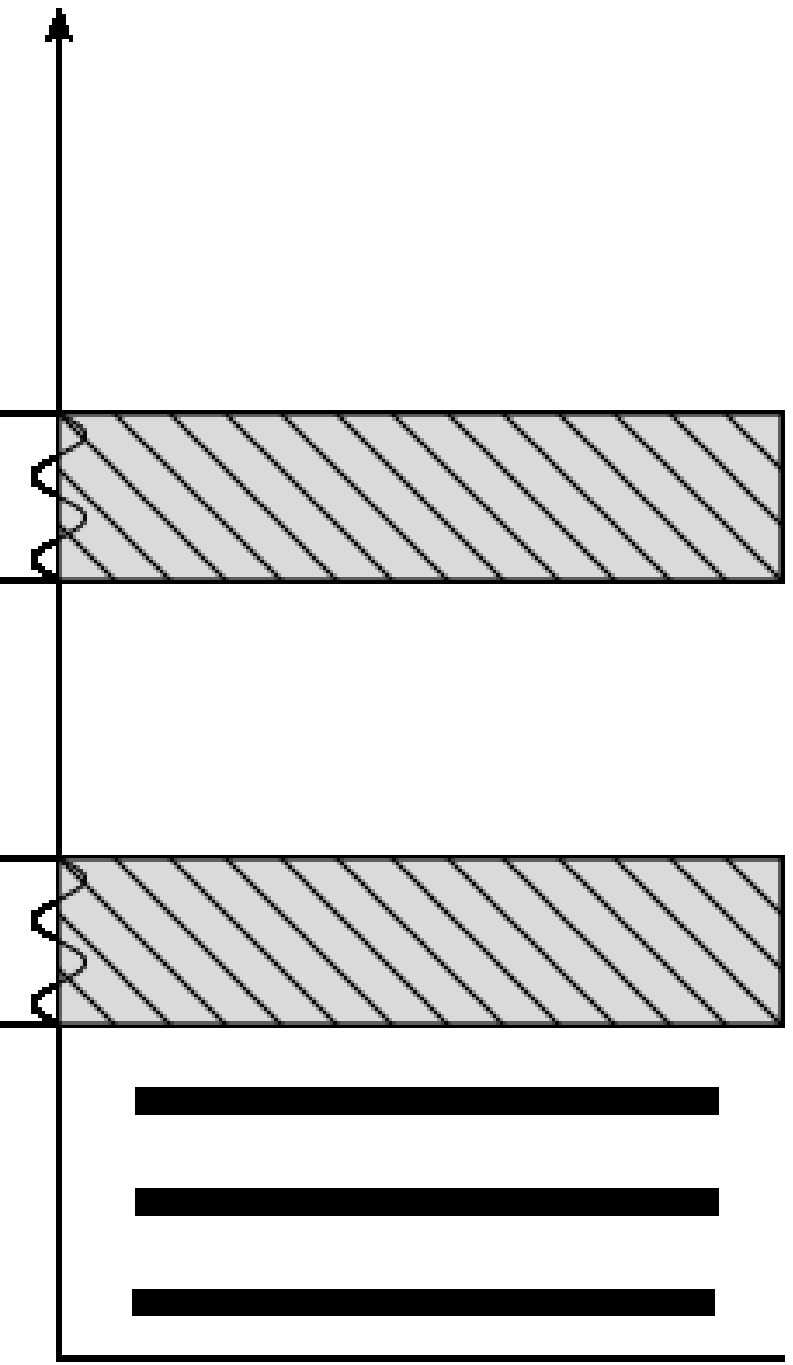}}}
\end{center}
\caption{The 10 RH neutrino mass patterns corresponding to
leptogenesis scenarios with
different sets of classical Boltzmann equations for the
calculation of the final asymmetry.}
\end{minipage}          
\end{figure}

\subsubsection{The (flavoured) $N_2$-dominated scenario}

Among the possible  10 RH neutrino mass patterns shown in Fig.~6,
for those three  where the $N_1$ wash-out occurs in 
the three fully flavoured regime, for $M_1 \ll 10^{9}\,{\rm GeV}$,
the asymmetry produced by the lightest RH neutrinos is much lower than
the observed value eq.~(\ref{etaBCMB}). The reason is that for these
low values of the RH neutrino masses, the flavoured $C\!P$ asymmetries
of the lightest RH neutrinos are too small. 

Therefore, the final asymmetry has necessarily to be produced by the 
heavier RH neutrinos, usually from the next-to-lightest RH neutrinos, 
since the $C\!P$ asymmetries of the
heaviest RH neutrinos $N_3$ are typically suppressed. 
The asymmetry can be produced from the $N_2$'s 
either  in the unflavoured (for $M_2 \gg 10^{12}\,{\rm GeV}$) or in 
the two fully flavoured  regime (for $10^{12}\,{\rm GeV} \gg M_2 \gg 10^9\,{\rm GeV}$).
These scenarios have some particularly attractive
features and correspond to a flavoured
$N_2$-dominated scenario \cite{vives} 
(it corresponds to the mass patterns (e), (f) and (g) in Fig.~6).  
  
While within the unflavoured approximation, assumed in the vanilla scenario,  
the wash-out from $N_1$
yields a global  exponential wash-out factor (cf. eq.~(\ref{N1washunfl})), when lepton flavour effects are 
taken into account  the asymmetry produced by the $N_2$'s, 
at the $N_1$ wash-out, distributes 
into an incoherent mixture of lepton flavour quantum eigenstates. 
It turns out that
the $N_1$ wash-out in one of the three flavours is negligible, corresponding to have
at least one $K_{1\a}\lesssim 1$,
in quite a wide region of the parameter space \cite{bounds}.
In this way, accounting for flavour effects, the region of applicability  of the
$N_2$-dominated scenario  enlarges considerably, since it is not
necessary that $N_1$ fully decouples but it is sufficient that it decouples
just in a specific lepton flavour. 

Recently, it has also been realised that,
accounting for the Higgs and for the quark asymmetries, the dynamics of the flavour asymmetries
couple and the lightest RH neutrino wash-out in a particular flavour can be circumvented even when $N_1$ is strongly
coupled in that flavour \cite{flcoupling}.
Another interesting effect arising in the $N_2$-dominated scenario is {\em phantom leptogenesis}.
This is a pure quantum-mechanical effect originating from the fact that
parts of the flavoured asymmetries, the phantom terms, escape completely
the wash-out at the production. These terms are proportional to the
initial $N_2$-abundance and, therefore, introduce an additional dependence
on the initial abundance of the RH neutrinos. 
More generally, it has been recently shown that phantom terms
associated to a RH neutrino species $N_i$ with $M_i \gg 10^{9}\,{\rm GeV}$
are non-vanishing if the initial $N_i$ abundance is non-vanishing. 
However, the phantom terms produced by the lightest RH neutrinos
cancel with each other and do not contribute to the final total asymmetry
though they can contribute to the flavoured asymmetries and could have 
potential applications, for example, in active-sterile neutrino oscillations
in the early Universe \cite{densitymatrix2}.

\subsubsection{Heavy neutrino flavour projection}

Even assuming a strong RH neutrino mass hierarchy, a coupled $N_1$ and 
$M_1\gtrsim 10^{12}\,{\rm GeV}$ (it corresponds to the first panel in Fig.~6), the asymmetry produced by the 
heavier RH neutrino decays,  in particular by the $N_2$'s decays,  
can be sizeable and in general is 
not completely washed-out by the lightest RH neutrino
processes. This is because there is in general an (orthogonal) component
that escapes the $N_1$ wash-out \cite{bcst} while the remaining
(parallel) component undergoes the  usual exponential wash-out. 
For a mild mass hierarchy, $\d_3 \lesssim 10$,
even the asymmetry produced by the $N_3$'s decays can be sizeable and circumvent
 the $N_1$ and $N_2$ wash-out. 
 
 Due to this effect of heavy neutrino flavour projection 
 and because of an additional contribution to the
 flavoured $C\!P$ asymmetries $\ve_{2\a}$,
 that is independent of the heaviest RH neutrino mass $M_3$ and that
 is not suppressed in the hierarchical limit,
 it has been recently noticed  that, when lepton flavour effects are consistently taken into account,
 the contribution from the next-to-lightest RH neutrinos can be dominant even
 in an effective  two RH neutrino model with $M_3\gtrsim 10^{15}\,{\rm GeV}$
 \cite{2RHNlep}.

\subsubsection{The problem of the initial conditions in flavoured leptogenesis}

As we have seen, within an unflavoured description, as in the vanilla scenario, the problem
of the dependence on the initial conditions reduces just to satisfy
the strong washout condition  
$K_1 \gg 1$. In other words there is a full correspondence between
strong wash-out and independence of the initial conditions. 

When (lepton and heavy neutrino) flavour effects are considered, the situation is much more involved and for example imposing  strong wash-out conditions on all flavoured
asymmetries ($K_{i\a}$) is not enough to guarantee independence of the initial conditions.
Perhaps the most striking consequence is that in a traditional 
$N_1$-dominated scenario there is no condition that can guarantee
 independence of the initial conditions.  The only possibility to have independence
 of the initial conditions is represented by a tauon $N_2$-dominated scenario 
 \cite{problem},
 that means a scenario where the asymmetry is dominantly produced
 from the next-to-lightest RH neutrinos, implying $M_2\gg 10^{9}\,{\rm GeV}$, 
 in the tauon flavour. The condition $M_1 \ll 10^{9}\,{\rm GeV}$ 
 is also important to have a projection of the quantum lepton states 
 on the orthonormal three lepton flavour
 basis prior to the lightest RH neutrino wash-out.
 
 \subsubsection{Density matrix formalism} 
  
 As in the case of the $N_1$-dominated scenario, a density matrix formalism
 allows to extend the calculation of the final asymmetry 
 when the RH neutrino masses fall in one of the two transition regimes
 between two fully flavoured regimes.  In this way, with a density matrix
 equation,  the calculation
 of the final asymmetry can be performed for a generic choice of the
 three RH neutrino masses. 
 Moreover within a density matrix formalism,  the validity
 of the multi-stage Boltzmann equations describing the ten asymptotic limits
 shown in Fig.~6, including  the presence of phantom terms and the flavour 
 projection effect, is fully confirmed \cite{densitymatrix2}. 

\section{Testing new physics with leptogenesis}

The see-saw mechanism extends the SM introducing eighteen new parameters
when three RH neutrinos are added. On the other hand, low energy
neutrino experiments can only potentially test nine parameters in the
neutrino mass matrix $m_{\nu}$. Nine high energy parameters, those characterising the properties
of the three RH neutrinos (three masses and six parameters in the see-saw orthogonal matrix $\O$ introduced 
in the eq.~(\ref{orthogonal}) encoding the three life times and the three total $C\!P$ asymmetries),
are not tested by low energy neutrino experiments.
Quite interestingly, the requirement of successful leptogenesis, 
\be
\eta_B^{\rm lep}(m_{\nu},\Omega,M_i)=\eta_{B}^{CMB}   \,  ,
\ee
provides an additional constraint on a combination
of both low energy neutrino parameters and high energy neutrino parameters.
However, just one additional constraint does not seem to be still sufficient to over-constraint the parameter
space leading to testable predictions. Despite this, as we have seen, in the vanilla leptogenesis scenario
there is an upper bound on the neutrino masses. The reason is that in this case
$\eta_B$ does not depend on the 6 parameters related to the properties of the two heavier RH neutrinos so that
the asymmetry depends on a reduced subset of high energy parameters (just three instead of nine). At the
same time, the final asymmetry gets strongly suppressed by the 
absolute neutrino mass scale when this is
larger than the atmospheric neutrino mass scale.
This is why imposing successful leptogenesis 
yields an upper bound on the neutrino masses.

When flavour effects are considered, the vanilla leptogenesis
scenario holds only under very special conditions. More generally
the parameters in the leptonic mixing matrix also
directly affect the final asymmetry and, accounting for flavour effects,
one could hope to derive definite predictions on the leptonic mixing matrix.
However, when flavour effects are taken into account,
the final asymmetry depends, in general, also on the 6 parameters associated to the two heavier RH neutrinos
that were canceling out in the calculation of the final asymmetry in the vanilla
scenario, at the expenses of predictability.
For this reason, in a general scenario with three RH neutrinos, it is not possible
to derive any prediction on low energy neutrino parameters.
An upper bound $m_i\lesssim 0.1$ on neutrino masses is still found but only
within the regime of applicability of the classical Boltzmann equations 
\cite{densitymatrix,bounds,josse}. This
could somehow get relaxed within a more general 
density matrix approach \cite{flavoreffects1,flavoreffects2,densitymatrix}.

In order to gain predictive power, two possibilities have been explored in the last years.
A first one is to consider non minimal scenarios giving rise to additional phenomenological constraints.
For example, we have already mentioned how with a non minimal see-saw mechanism it is possible to lower
the leptogenesis scale and have signatures at colliders. It has also been noticed that
in supersymmetric models one can enhance the branching ratios of lepton flavour violating processes
or electric dipole moments and in this way the existing experimental bounds
further constrain the see-saw parameter space \cite{lfvedm}.
A second possibility is to search again, as for vanilla leptogenesis, for a reasonable
scenario where the final asymmetry depends on a reduced number of independent parameters
over-constrained by the successful leptogenesis bound. Let us briefly discuss some
of the main ideas that have been proposed in this second respect.

\subsection{Two RH neutrino model}

A  well motivated scenario that attracted great attention is a two
RH neutrino model \cite{2RHN}, where the third RH neutrino is either absent or
effectively decoupled from the see-saw mechanism. 
This necessarily happens when $M_3\gg 10^{14}\,{\rm GeV}$, implying that the
lightest left-handed neutrino mass $m_1$ has to vanish. It can be shown that the number of parameters
gets reduced from 18 to 11.

The two RH neutrino model has been traditionally considered as a sort of benchmark case
for the $N_1$-dominated scenario, where the final asymmetry is dominated by the contribution
from the lightest RH neutrinos. However recently \cite{2RHNlep}, as we anticipated already, it has been
shown that there are some regions in the high energy parameter space (basically one complex angle)  that are $N_2$-dominated.
Even though in this model there is still a lower bound on $M_1$, contrarily to the $N_2$
dominated scenario with three RH neutrinos, in these new $N_2$-dominated regions the lower bound
on $M_1$ is relaxed by one order of magnitude, down to about $10^{11}\,{\rm GeV}$. Interestingly these regions
correspond to so called light sequential dominated models \cite{sequential}. It should be said that in
a general 2 RH neutrino model it is still not possible to make predictions on the low energy neutrino
parameters. To this extent one should further reduce the parameter space for example assuming texture zeros
in the neutrino Dirac mass matrix.

\subsection{$SO(10)$ inspired models}

 In order to gain  predictive power, one can
impose conditions within some model of new physics embedding the see-saw mechanism.
An interesting example is represented by the `$SO(10)$-inspired scenario' \cite{branco},
where $SO(10)$-inspired conditions are imposed on the neutrino Dirac mass matrix $m_D$.
In the basis where the charged leptons mass matrix and the Majorana mass matrix are diagonal,
one can write in the bi-unitary parametrisation  $m_D = V_L^{\dagger}\,D_{m_D}\,U_R$,
where $D_{m_D}\equiv {\rm diag}({\l_1,\l_2,\l_3})$ is the diagonalised neutrino Dirac mass matrix.
In $SO(10)$-inspired models, the unitary matrix $V_L$ is related and not too different
from the so called Cabibbo-Kabayashi-Maskawa quark mixing matrix $V_{CKM}$.
The $U_R$ can then be calculated from $V_L$, $U$ and $m_i$,
considering that, as it can be seen from the see-saw formula eq.~(\ref{see-saw}),
it provides a Takagi factorisation of
$M^{-1} \equiv D^{-1}_{m_D}\,V_L\,U\,D_m\,U^T\,V_L^T\,D^{-1}_{m_D}$,
or explicitly $M^{-1} = U_R\,D_M^{-1}\,U_R^T$. Moreover the three eigenvalues
$\l_1,\l_2,\l_3$ are assumed to be order-of-magnitude wise equal to the three up quark masses. 

In this way the RH neutrino masses and the matrix $U_R$ are expressed in terms of the
low energy neutrino parameters, of the eigenvalues $\l_i$ of $m_D$ and of the parameters in $V_L$.
Since one has typically $M_1 \sim 10^{5}\,{\rm GeV}$ and $M_{2}\sim 10^{11}\,{\rm GeV}$,
the asymmetry produced from the lightest RH neutrino decays is negligible and the
$N_2$-dominated scenario is realised \cite{SO10}. Imposing the leptogenesis bound
and considering that the final asymmetry does not depend on $\l_1$ and on $\l_3$, one obtains
constraints on all low energy neutrino parameters slightly depending just on the
parameter $\l_2$ typically parameterised in terms of $\alpha_2\equiv \l_2/m_c$, where $m_c$
is the charm quark mass. Some examples of the constraints on the low energy neutrino parameters
are shown in Fig.~7
for a scan over the $2\sigma$ ranges of the allowed values of the
low energy parameters and over the parameters
in $V_L$ assumed to be $I< V_L < V_{CKM}$ \cite{SO10} and
for three values of $\alpha_2=5,4,\alpha_{\rm min}\simeq 1$.
This scenario has been also extended
including a type II contribution to the see-saw mechanism 
usually arising  in left-right symmetric models \cite{abada}.
\begin{figure}
\begin{center}
\leftline{\hfill\vbox{\hrule width 5cm height0.001pt}\hfill}
         \mbox{\epsfig{figure=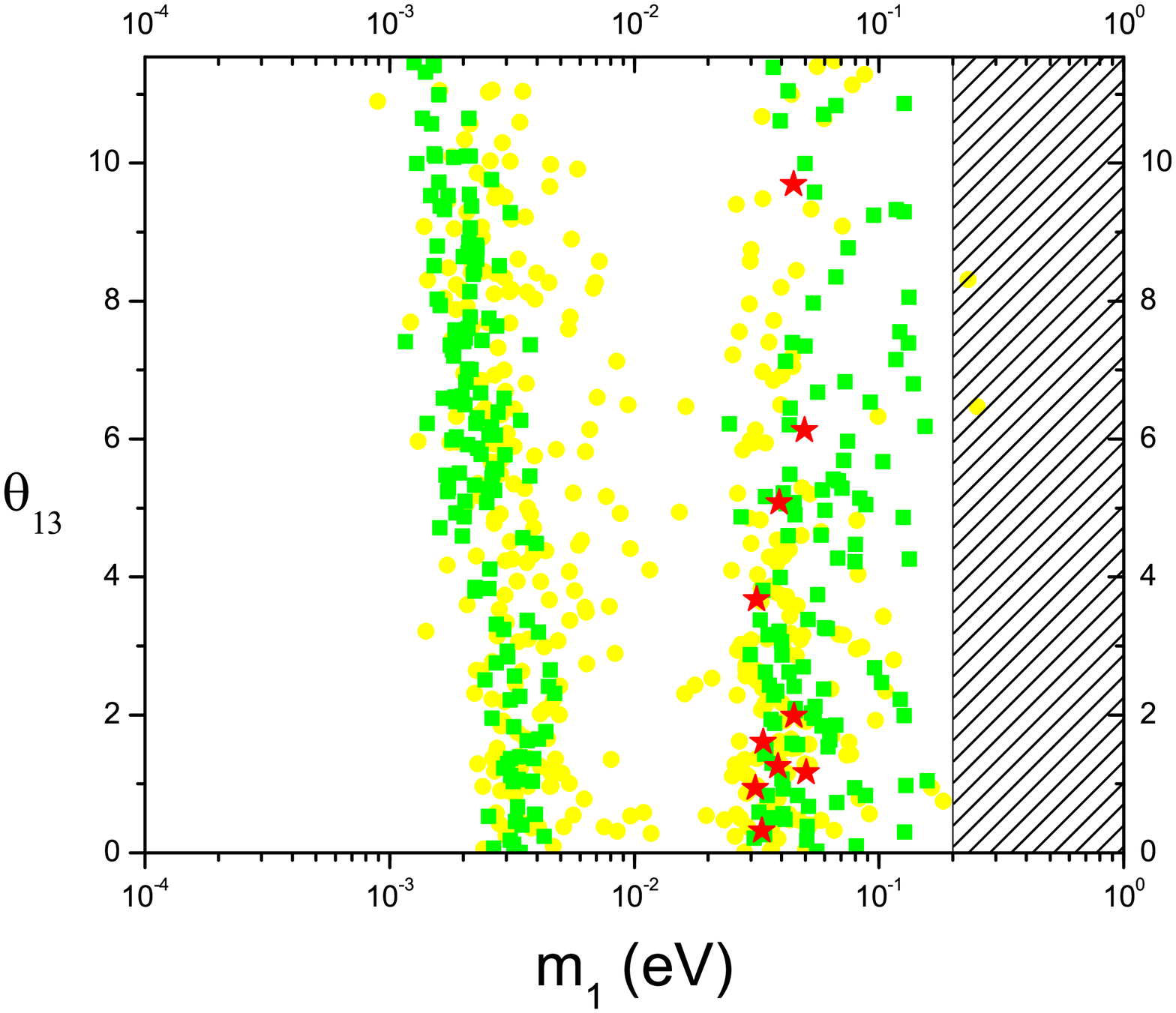,width=3.5cm,height=3.5cm}}
         \mbox{\epsfig{figure=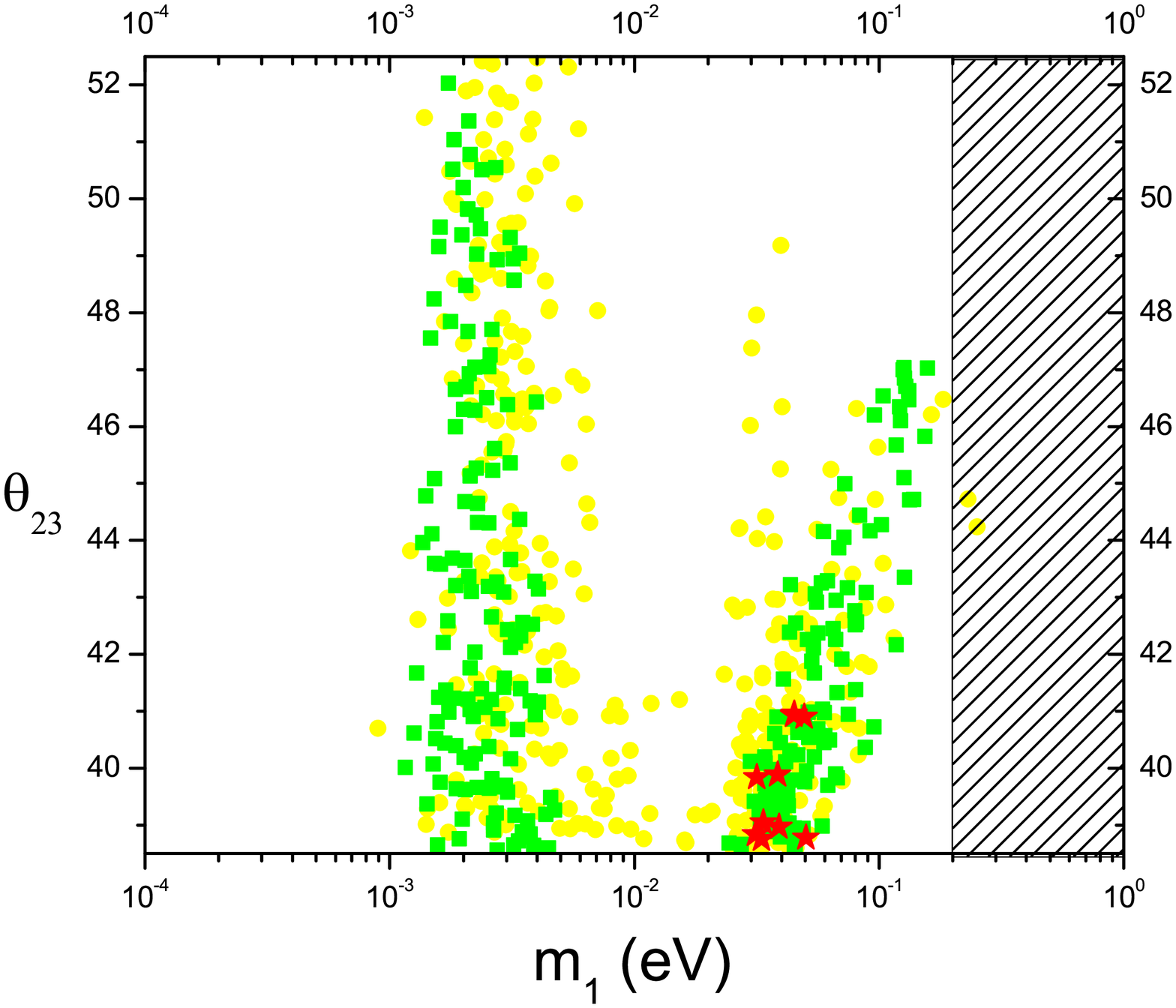,width=3.5cm,height=3.5cm}} \\
         \mbox{\epsfig{figure=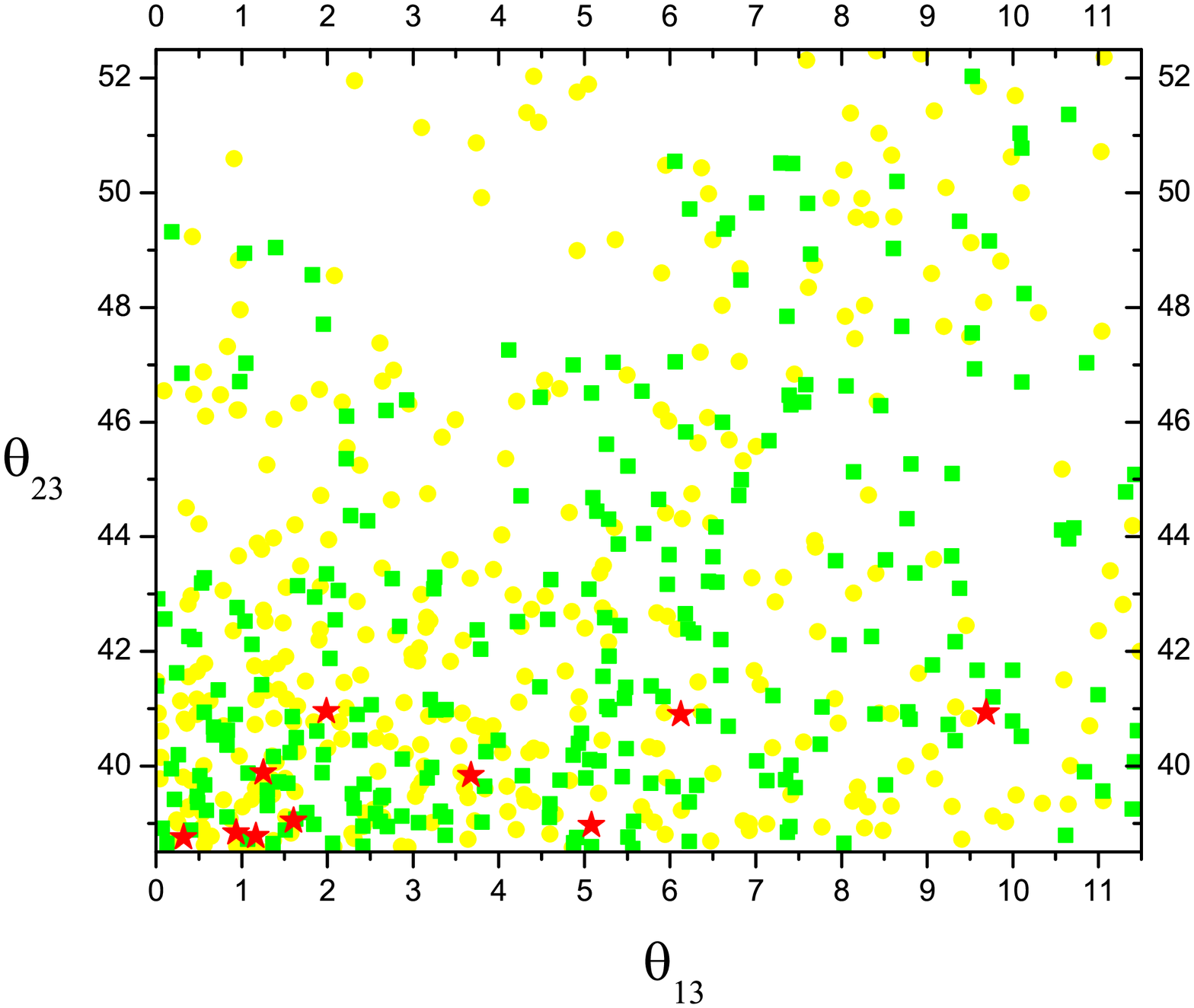,width=3.5cm,height=3.5cm}}
         \mbox{\epsfig{figure=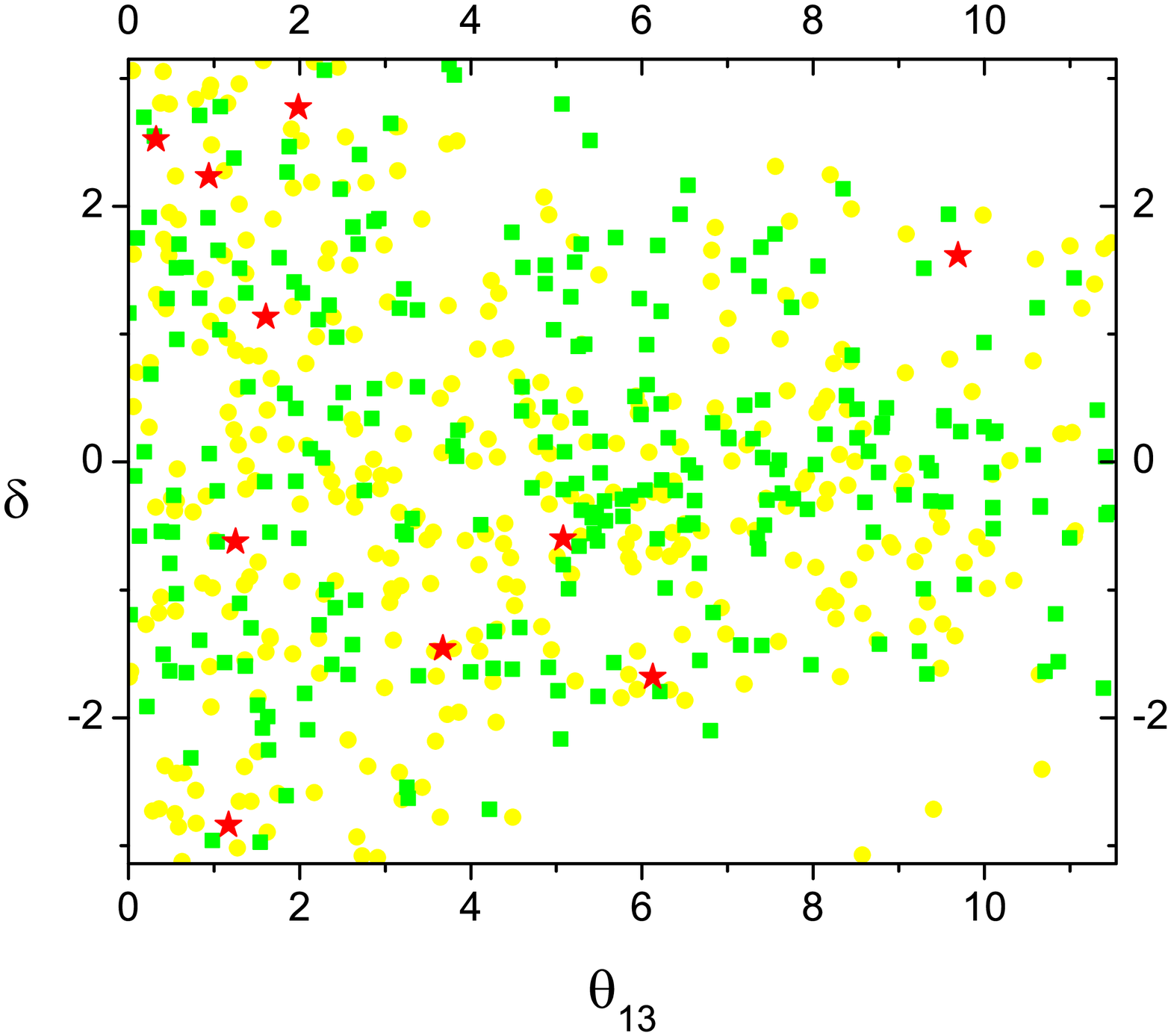,width=3.5cm,height=3.5cm}}
\leftline{\hfill\vbox{\hrule width 5cm height0.001pt}\hfill}
\caption{Constraints on some of the low energy neutrino parameters in the $SO(10)$-inspired
         scenario for normal ordering and $I< V_L < V_{CKM}$  \cite{SO10}. The yellow, green and red points
         correspond respectively to $\a_2=5,4,1$.}
\end{center}
\end{figure}

\subsection{Discrete flavour symmetries}

 An account of heavy neutrino flavour effects is important when leptogenesis is embedded within
theories that try to explain  tribimaximal mixing  for the leptonic
mixing matrix via flavour symmetries. It has been shown in particular that
if the symmetry is unbroken then the $C\!P$ asymmetries of the RH neutrinos would exactly
vanish since the basis of the three heavy neutrino flavours would be exactly orthonormal. 
On the other hand when the symmetry is broken, for the naturally expected
values of the symmetry  breaking parameters, then the observed
matter-antimatter asymmetry can be successfully reproduced \cite{feruglio}.
It is interesting that in a minimal picture based on a $A4$ symmetry
\setcounter{footnote}{18}\footnote{A4 is the group of the even permutations of four objects, so it has 12 elements.
From a geometrical point of view, it is the subgroup of the 12 possible three-dimensional rotations
leaving invariant a regular tetrahedron and in this case it is a symmetry for this system. In our context
an $A4$-symmetry  is then equivalent to impose that the Lagrangian 
of the new model extending the SM, 
is invariant under transformations of the $A4$ group acting on the flavour space.}
, one has a RH neutrino mass spectrum with
$10^{15}\,{\rm GeV} \gtrsim M_3 \gtrsim M_2 \gtrsim  M_1 \gg 10^{12}\,{\rm GeV}$. One has therefore
that all the asymmetry is produced in the unflavoured regime and that the mass spectrum
is only mildly hierarchical (it has actually the same kind of hierarchy of light neutrinos).
At the same time the small symmetry breaking implies corresponds to
quasi-orthonormality of the three lepton quantum states produced in the RH neutrino
decays. Under these conditions the wash-out of the asymmetry produced by one RH neutrino species
from the inverse decays of a lighter RH neutrino species is essentially negligible. The final
asymmetry then receives a non negligible contribution from the decays of all three RH neutrinos species.

\subsection{Supersymmetric models}

Within a supersymmetric framework the final asymmetry calculated in
the vanilla leptogenesis scenario undergoes  minor changes \cite{proceedings}.
However, supersymmetry introduces a conceptual important issue: the stringent
lower bound on the reheat temperature, $T_{\rm RH}\gtrsim 10^{9}\,{\rm GeV}$,
is typically marginally compatible with an upper bound
from the avoidance of the so called gravitino problem $T_{\rm RH}\lesssim 10^{6-10}\,{\rm GeV}$, with the
exact value depending on the parameters of the model \cite{gravitino}  
\setcounter{footnote}{19}\footnote{The gravitino problem is the observation
that within a certain popular class of supersymmetric models, so called gravity mediated supersymmetric
models, for large reheating temperatures, $T_{\rm RH}\gtrsim 10^{6-10}\,{\rm GeV}$ (the exact value depending
on many of the new supersymmetric parameters) the supersymmetric partners of gravitons,
so called gravitinos, would be produced in a way either to yield a matter energy density parameter
$\O_{M,0}$ much above the observed value or to spoil the predictions of standard Big Bang Nucleosynthesis,
the model that, within the Hot Big Bang models, successfully reproduces the measured values 
of the primordial nuclear abundances. For this reason, within these models, an upper bound
$T_{\rm RH}\lesssim 10^{6-10}\,{\rm GeV}$ has to be imposed.}.
It is quite remarkable
that the solution of such a issue inspired an intense research activity on supersymmetric
models able to reconcile minimal leptogenesis and the gravitino problem. Of course on the
leptogenesis side, some of the discussed extensions beyond the vanilla scenario that relax the neutrino
mass bounds also relax the $T_{\rm RH}$ lower bound. However, notice that in the $N_2$ dominated
scenario, while the lower bound on $M_1$ simply disappears, there is still a lower bound
on $T_{\rm RH}$ that is even more stringent, $T_{\rm RH}\gtrsim 6\times 10^{9}\,{\rm GeV}$ \cite{geometry}.
As mentioned already, with flavour effects one has the possibility to relax the lower bound
on $T_{\rm RH}$ if a mild hierarchy in the RH neutrino masses
is allowed together with a mild cancellation in the see-saw formula \cite{bounds}.
However for most models, such as sequential dominated models \cite{sequential},
this solution does not apply. 

A major modification introduced by supersymmetry
is that the critical value of the mass of the decaying RH neutrinos
setting the transition from an unflavoured regime to a two-flavour regime
and from a two-flavour regime to a three flavour regime is enhanced by a factor $\tan^2\beta$ \cite{flavoreffects1,antusch}.
This has a practical relevance in the calculation of the asymmetry within supersymmetric models
and it is quite interesting that leptogenesis becomes sensitive to such a relevant
supersymmetric parameter. 

\section{A wish list for leptogenesis}

In 2011 two important  experimental results have been announced that,
if confirmed, can be interpreted as quite important toward 
the possibility  of  a positive test of  leptogenesis 
\setcounter{footnote}{20}\footnote{Having occurred in the 25th anniversary of leptogenesis \cite{fy}, 
they could quaintly be regarded as a sort of birthday present.}. 
The first result is the strong hint of a non-vanishing 
$\theta_{13}\simeq 9^{\circ}$ from T2K results first and then
confirmed by many other experiments at more than $5\sigma$ standard deviations \cite{theta13}.
Very interestingly, such a  value is sufficiently large to  open 
realistically the possibility of a measurement of 
the neutrino oscillation $C\!P$ violating invariant
$J_{CP}\propto \sin\theta_{13}\,\sin\delta$ during next years. 

If this measurement will give a non-vanishing result, this will have some direct model independent important consequences.
First of all, even the small contribution to the final asymmetry uniquely stemming 
from a Dirac phase could be sufficient to reproduce the observed final
asymmetry \cite{flavorlep,bounds}.  In any case, the presence of $C\!P$ violation at low energies would
certainly support also the presence of $C\!P$ violation at high energies, since
given a generic theoretical model predicting the neutrino Dirac mass matrix $m_D$,
these are in general both present. However, there is an even more important
practical relevance of such a measurement, even if in the end this will
indicate a vanishing value of $J_{CP}$ within the experimental error. 
The reason is that such a measurement of $J_{CP}$
will provide an additional constraint in specific see-saw models able to produce
simultaneous predictions on the low energy neutrino parameters and on
the final asymmetry in leptogenesis.  The interesting case of $SO(10)$-inspired
leptogenesis offers a definite example.  By plugging more and more
information from low energy neutrino experiments, we will be able 
more and more strongly to test the models.

At the same time the  hint of a $\simeq 125\,{\rm GeV}$ Higgs boson
reported by the ATLAS and CMS collaborations can be also interpreted
as a positive result for leptogenesis. First of all because leptogenesis 
relies on the Higgs mechanism and second because the measurement
of the Higgs boson mass could in future  open new opportunities for additional
phenomenological information to be imposed on leptogenesis scenarios
for example relying on the requirement of vacuum stability \cite{giudiceetal}. 

What are other possible future experimental developments 
that could be important for testing leptogenesis? 
An improved information from absolute neutrino mass scale experiments,
both on the sum of the neutrino masses from cosmology and on $m_{ee}$
from $0\n\b\b$ experiments could be crucial especially if these
phenomenologies will definitively confirm the upper bound $m_1 \lesssim 0.1\,{\rm eV}$
but at the same time will detect a positive signal $m_1 \sim 10^{-2}\,{\rm eV}$,
within the reach of future experiments in next 5-10 years. In this case  
the a desirable leptogenesis neutrino mass window   
$10^{-3}\,{\rm eV}\lesssim m_1 \lesssim 10^{-1}\,{\rm eV}$ 
would be confirmed 
\setcounter{footnote}{21}
\footnote{This window was found to be optimal within the vanilla scenario \cite{upperbound}. 
However, we discussed how flavour effects can
potentially relax the upper bound but solutions for $m_1\gtrsim 0.1\,{\rm eV}$ 
still need to be confirmed within a density matrix formalism and, therefore, 
a result $m_1\lesssim 0.1\,{\rm eV}$ would be, if not a necessity, a safe finding and certainly desirable. 
For $m_1\gtrsim 10^{-3}\,{\rm eV}$ the strong thermal condition (independence of the initial conditions) 
would necessarily hold in the vanilla scenario. Even though, including flavour effects additional conditions 
need to be imposed to realise strong thermal leptogenesis, still, if
$m_1\gtrsim 10^{-3}\,{\rm eV}$ will be found, it would be much more natural to satisfy them.
For these reasons  the window $10^{-3}\,{\rm eV}\lesssim m_1 \lesssim 10^{-1}\,{\rm eV}$
can still be regarded as a desirable window for the absolute neutrino mass scale in leptogenesis.}.
At the same time it is likely that a measurement of the neutrino mass ordering 
will be possible when the information from absolute neutrino mass scale experiments is combined
with neutrino oscillation experiments. Moreover, 
with a large $\theta_{13} \sim 9^{\circ}$ value, even just neutrino oscillation experiments 
could  be sufficient to determine the mass ordering within next years. In this case 
a precise measurement of the allowed region in the plane $m_{ee}-\sum_i\,m_i$
would provide a positive test of specific leptogenesis scenarios, such as 
$SO(10)$-inspired leptogenesis, especially when a  condition of independence
of the initial conditions is imposed \cite{prep}. 
It is therefore quite exciting that even just the expected
new information from (in most cases 
already data taking) low energy neutrino experiments will be able to 
test models of leptogenesis in a substantial way during next years. 
In conclusion, with leptogenesis, we have the fascinating opportunity to probe at the same time
both a new stage in the early Universe history in cosmology and physics beyond the SM.

\subsection{Acknowledgements}

PDB acknowledges financial support 
from the NExT/SEPnet Institute, 
from the STFC Rolling Grant ST/G000557/1 and
from the  EU FP7  ITN INVISIBLES (Marie Curie Actions, PITN- GA-2011- 289442).
I wish to thank S.~Antusch, E.~Bertuzzo, S.~Blanchet, W.~Buchmuller, F.~Feruglio, D.~Jones,
S.~King, L.~Marzola, M.~Plumacher, G.~Raffelt, A.~Riotto for a fruitful collaboration
on leptogenesis.  I also wish to thank Asmaa Abada, Steve Blanchet and Wilfried Buchm\"{u}ller for useful comments and the  
Laboratoire de Physique Th\'{e}orique, Universit\'{e} de Paris-Sud 11 (Orsay) 
for the warm hospitality during the completion of this work.
This review is dedicated in memory of Alexey Anisimov.

%
 

\end{document}